\documentclass[iop]{emulateapj}
\usepackage{txfonts}
\usepackage{graphicx}
\usepackage{natbib}
\accepted{13 Feb 2012}

\shorttitle{Jet rotation investigated in the near-ultraviolet with HST/STIS}
\shortauthors{Coffey et al. 2012}

\begin{document}

\title{Jet rotation investigated in the near-ultraviolet with HST/STIS}

\author{Deirdre Coffey\altaffilmark{1, 2} \email{dac@cp.dias.ie}}
\author{Elisabetta Rigliaco\altaffilmark{1}}
\author{Francesca Bacciotti\altaffilmark{1}\email{fran@arcetri.astro.it}}
\author{Thomas P. Ray\altaffilmark{2}}
\author{Jochen Eisl\"{o}ffel\altaffilmark{3}}

\altaffiltext{1}{Osservatorio Astrofisico di Arcetri, 50125 Firenze, Italy}
\altaffiltext{2}{The Dublin Institute for Advanced Studies, Ireland}
\altaffiltext{3}{Th\"{u}ringer Landessternwarte Tautenburg, Germany}

\begin{abstract}

We present results of the second phase of our near-ultraviolet investigation into protostellar jet rotation using HST/STIS. We obtain long-slit spectra at the base of five T~Tauri jets to determine if there is a difference in radial velocity between the jet borders which may be interpreted as a rotation signature. These observations are extremely challenging and push the limits of current instrumentation, but have the potential to provide long-awaited observational support for the magneto-centrifugal mechanism of jet launching in which jets remove angular momentum from protostellar systems. We successfully detect all five jet targets (from RW~Aur, HN~Tau, DP~Tau and CW~Tau) in several near-ultraviolet emission lines, including the strong Mg~II doublet. However, only RW Aur's bipolar jet presents sufficient signal-to-noise for analysis. The approaching jet lobe shows a difference of 10~km~s$^{-1}$ in a direction which agrees with the disk rotation sense, but is opposite to previously published optical measurements for the receding jet. The near-ultraviolet difference is not found six months later, nor is it found in the fainter receding jet. Overall, in the case of RW~Aur, differences are not consistent with a simple jet rotation interpretation. Indeed, given the renowned complexity and variability of this system, it now seems likely that any rotation signature is confused by other influences, with the inevitable conclusion that RW~Aur is not suited to a jet rotation study. 
\\ 

\end{abstract}

\keywords{ISM: jets and outflows --- stars: formation,  --- 
stars: individual (RW Aur, HN Tau, DP Tau, CW Tau)  --- stars: pre-main sequence}

\section{Introduction}
\label{introduction}

Young stars must loose angular momentum in order to accrete gas from their circumstellar disks and continue to grow to their final pre-main sequence mass. However, difficulties remain in understanding the mode by which the angular momentum is extracted from protostellar systems. It is proposed that, during the accretion process, magneto-centrifugal forces are responsible for the launch, acceleration and collimation of high velocity protostellar jets (\citealp{Pudritz2007}; \citealp{Shang2007}). Long standing observational obstacles in testing proposed models lie in the fact that young stars are often heavily embedded, infall and outflow kinematics are complex and confused close to the source, and the spatial scales are relatively small \citep{Ray2007}. 

With the advent of high resolution ($<$1$\arcsec$) observations, there have been the first reports of detections of rotation signatures in jets from young stars. These measurements of differences in radial velocity across the jet seem to constitute observational indications that jets are indeed the main player in transporting angular momentum from protostellar systems. Initial tentative claims from ground-based data \citep{Davis2000} were followed by high resolution measurements close to the source in Hubble Space Telescope Imaging Spectrograph (HST/STIS) optical spectra (\citealp{Bacciotti2002}; \citealp{Woitas2005}). Follow-up survey observations with HST/STIS in the optical regime confirmed that systematic radial velocity asymmetries across the jet base are common in T Tauri systems \citep{Coffey2004}. The survey was moved to the higher spatial and spectral resolution afforded by the near-ultraviolet (NUV), and yet again asymmetries were identified in two of the original six T Tauri star sample \citep{Coffey2007}. 

These findings have profound implications as they may represent the long-awaited observational support for 
the magneto-centrifugal class of models. Indeed, the derived toroidal velocities are in 
agreement with model predictions. Furthermore, these findings may also potentially act as a powerful 
discriminant between competing steady MHD models (\citealp{Bacciotti2002}; \citealp{Anderson2003}; \citealp{Coffey2004}; 2007; \citealp{Ferreira2006}). Due to these implications, the results have triggered much debate 
as to whether the velocity asymmetries should indeed be interpreted as a jet rotation 
signature. Alternative explanations include asymmetric shocking, jet precession \citep{Cerqueira2006} and entrainment by a warped disk \citep{Soker2005}. Indeed, in one case, RW Aur, the sense of the jet gradient in optical lines (\citealp{Coffey2004}; \citealp{Woitas2005}) does not 
match that of the disk \citep{Cabrit2006}, thus calling into question the jet rotation interpretation. 
However, this is a complex and highly variable multiple system and, while there are no signs of jet precession, it should be 
examined in greater detail to determine whether other influences are coming into play. 

Following the failure of the HST/STIS power supply in August 2005, 
which interrupted our original NUV survey, our efforts continued 
from the ground. Seeing limited near-infrared observations failed to resolve the atomic jet width 
(\citealp{Coffey2010}; 2011). (Use of adaptive optics was not possible due to the lack of suitable guide stars.) 
Although the molecular flow was resolved transversely in younger sources, and 
gradients detected, \citep{Coffey2010} the necessity of observing far from these embedded 
sources (hundreds to thousands of AU) introduces uncertainty as to whether any rotation signature will maintain its integrity to 
such distances. Meanwhile, several ground based findings for molecular flows were reported by other groups (e.g. \citealp{Codella2007}; \citealp{Lee2007}; \citealp{Chrysostomou2008}; \citealp{Correia2009}), including CB 26 \citep{Launhardt2009} and NGC 1333 IRAS 4A2 (\citealp{Choi2010}; 2011) both of which show agreement in jet and disk rotation sense. Other reports claim no gradients detected in molecular lines (e.g. \citealp{Guilloteau2008}). However, since the molecular flow traces the outer regions of the jet, the rotation is expected to be lower than the central atomic component and thus may fall within resolution constraints. 

With the repair of HST/STIS, we were successful in our bid for time to complete our HST/STIS NUV survey. The NUV \ion{Mg}{2} $\lambda\lambda$2796,2803 doublet offers roughly ten times more flux than the more commonly observed optical forbidden emission lines. Furthermore, HST/STIS spectral and spatial resolution doubles when moving from the optical to the NUV regime. Therefore, we can access the more collimated jet core which is traced by the higher excitation NUV lines. Observations of five jet targets have been conducted, and include the bipolar jet from RW Aur. 

\section{Observations}
\label{observations}

HST/STIS NUV observations were conducted of five T\,Tauri jet targets to ascertain if signatures of jet rotation are identifiable (cycle 17, proposal ID 11660). The target list comprised both lobes of the bipolar jet from RW\,Aur, the approaching jet lobes from CW\,Tau and HN\,Tau, and the receding jet lobe from DP\,Tau. The targets were originally chosen because they are nearby (140 pc), they emit strongly at optical wavelengths close to the star ($<$1$\arcsec$), their position angles are known, and the jet radial velocity are high enough to avoid signiÞcant interstellar or self-absorption at low blue-shifted velocities since \ion{Mg}{2} arises from permitted transitions. If the jet radial velocity is too low, the intensity peak is disrupted by absorption and so cannot be accurately fitted. 

A single long slit was placed perpendicular to the jet with an offset from the star of 0$\farcs$3 for most targets, with the exception of CW\,Tau and RW Aur approaching jets where the offset was  0$\farcs$2. For all targets, located in Taurus-Auriga at $\sim$140\,pc, the offset corresponds to a deprojected distance along the jet of less than 100 AU. The NUV MAMA detector was used with the E230M echelle grating, centered on 2707 \AA, and with a long slit of aperture 6 $\times$ 0.2 arcsec$^{2}$ to ensure the full width of the jet was observed. Wavelength coverage was from 2280 \AA\, to 3120 \AA, which allows detection of the shock-excited NUV double \ion{Mg}{2} at vacuum wavelengths 2796.352 and 2803.531 \AA. The nominal average dispersion per pixel was $\lambda$/60,000. The spectral sampling was 0.045 \AA\,pixel$^{-1}$, corresponding to a radial velocity sampling of 5 km s$^{-1}$ pixel$^{-1}$. Allowing for the degradation in velocity centroid precision, which inevitably arises from use of the E230M grating with the long-slit on an extended source, we achieve a velocity resolution of $\sim$30\,km\,s$^{-1}$. Using profile fitting, the effective velocity resolution achieved is heavily signal-to-noise (S/N) dependent, but typically 1$\sigma$ of 5 km s$^{-1}$ for S/N $>$ 20 \citep{Agra-Amboage2009}. The spatial sampling was 0.029 pixel$^{-1}$, while the spatial resolution was twice the sampling (i.e., a two pixel resolution of 0$\farcs$058) and the PSF has a FWHM of 0$\farcs$05 \citep{Robinson97}. Observation dates and exposure times are detailed in Table~\ref{exposure_times}. Essentially, two or three long exposures of typically 2500\,s were obtained for each jet target. Observations with an anti-parallel slit were also obtained for both lobes of the RW\,Aur bipolar jet, to ensure we could rule out instrumental effects. 

\section{Data Recalibration}
\label{recalib}

Although the data were largely processed through the standard HST pipeline, the combination of an echelle grating with a long slit means that wavelength calibration was not conducted automatically as part of the pipeline procedure. This is because a projection of the long slit on the detector can cause overlap of adjacent echelle orders, which may lead to errors in the automated pipeline wavelength calibration. However, this is not an issue in our case since the jet is not wide enough to result in order overlap. In addition to the necessity to carry out wavelength calibration, the delivery of two revised calibration reference files meant that a full recalibration of the data was required (see Section 3.5 of the HST STIS Data Handbook 6.0, May 2011). 

\subsection{Reference files}
\label{ref_files}

To facilitate wavelength calibration, lamp exposures were obtained automatically (AUTO wavecal), with the same slit as the science exposures (i.e. the 6$\times$0$\farcs$2 long slit). In order to ensure accurate zeropoint wavelength calibration, we requested that an additional lamp exposure with the 0$\farcs$2$\times$0$\farcs$2 slit (GO wavecal) be taken during the observations after each science exposure. We replaced the AUTO wavecal with the GO wavecal as the wavelength calibration reference file. 

Further replacements of calibration reference files related to dark current and flux calibration. Due to an unexpected $\sim$30$\%$ increase in the MAMA dark current after {\it HST} Service Mission 4 (SM4), an updated Temperature Dependant Characteristics (tdc) reference file was delivered in January 2011. Also, identification of corruption in the reference file for flux calibration led to delivery of an updated Time Dependent Sensitivity (tds) table reference file in November 2010, promising accuracy in flux detections to 6\% for the STIS MAMA E320M mode. 

Once the raw science file header keywords (wavcal, tdctab and tdstab) were updated to the correct reference files, the relevant calibration switch was updated so that wavelength calibration was also conducted as part of the recalibration, i.e. the keyword x2dcorr was set to PREFORM. 

\subsection{Recalibration procedure}

Recalibration was conducted by running the {\it CALSTIS} tool in PyRAF. Unfortunately, calstis cannot fully calibrate the data because it is not equipped to correct for the known tilt and curvature of the long slit image on the MAMA detector (see Sections 4.3.2 and 12.2 of the STIS Instrument Handbook). We overcome this by making use of our long slit AUTO wavcal. In our case, the best calibration of the data was achieved as follows. The Aperture Description (apd) table reference file was edited such that the angle column in row 32 read zero rather than 0.27 degrees (to avoid too many resampling steps). Using the raw GO wavcal as the wavelength calibration reference file in both cases, calstis was run on the raw science exposure {\it and} the raw AUTO wavecal exposure (treating it as though it were a science exposure). The outputs (file extension x2d) for the science exposures were co-added at this point. The \ion{Mg}{2} doublet emission (order 73) can be found in extension 22. Extension 22 of the AUTO wavecal (x2d) file was then analysed for tilt and curvature (via IRAF's identify, reidentify, fitcoords routines), and the resulting correction was applied to extension 22 of the co-added science file (via IRAF's transform routine). Finally, any reflected continuum emission was subtracted (though it was very faint in all cases). 

\section{Data analysis and results}
\label{results}

\begin{figure}
\begin{center}
\epsscale{1.0}
\plotone{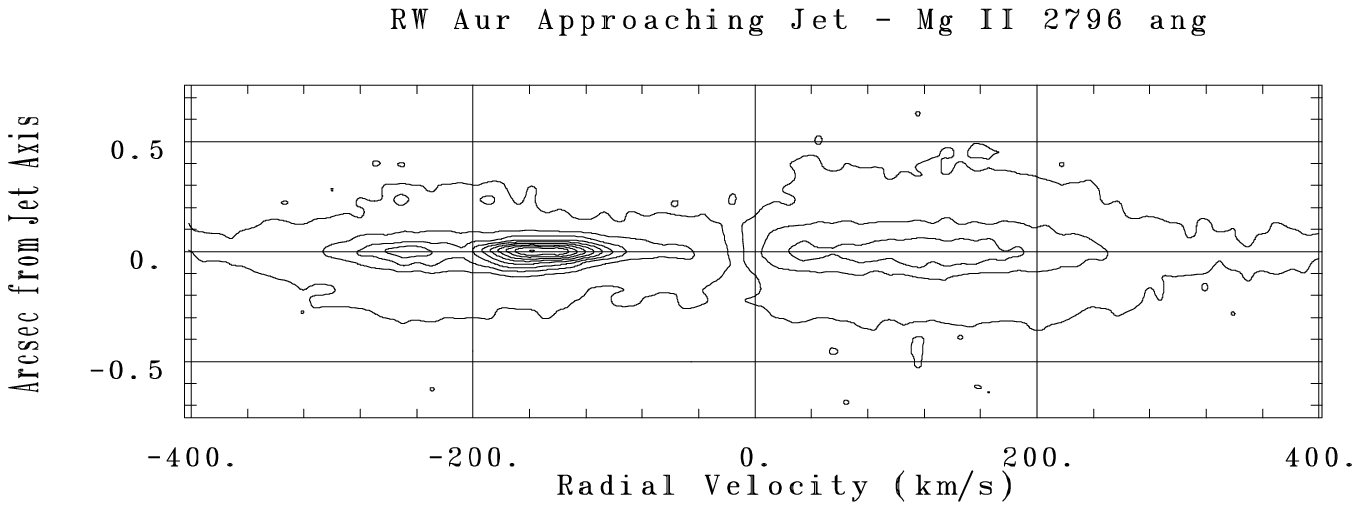} 
\plotone{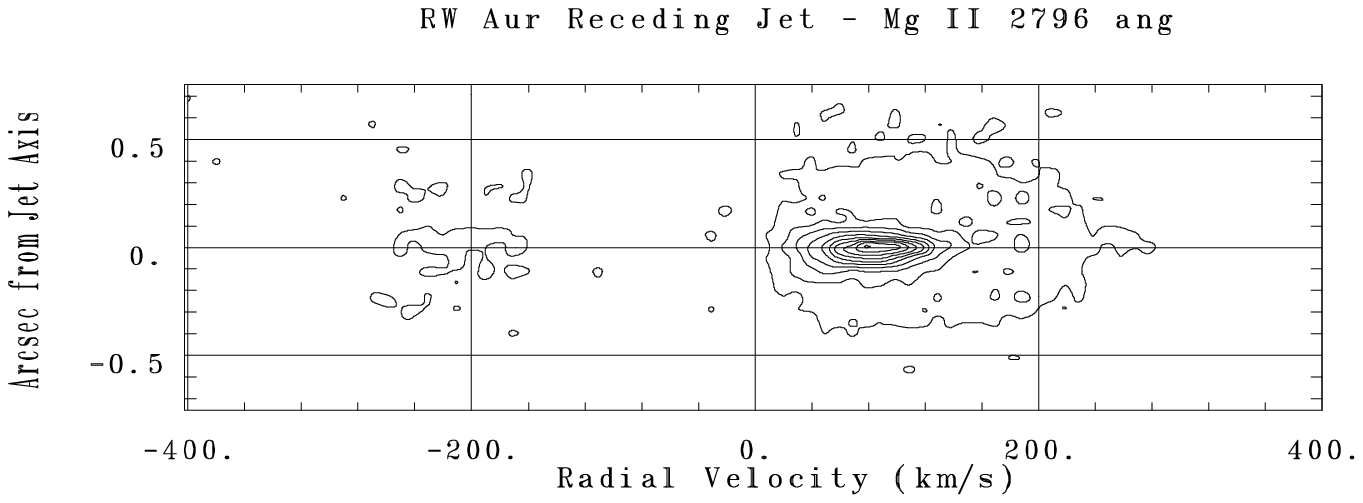}
\plotone{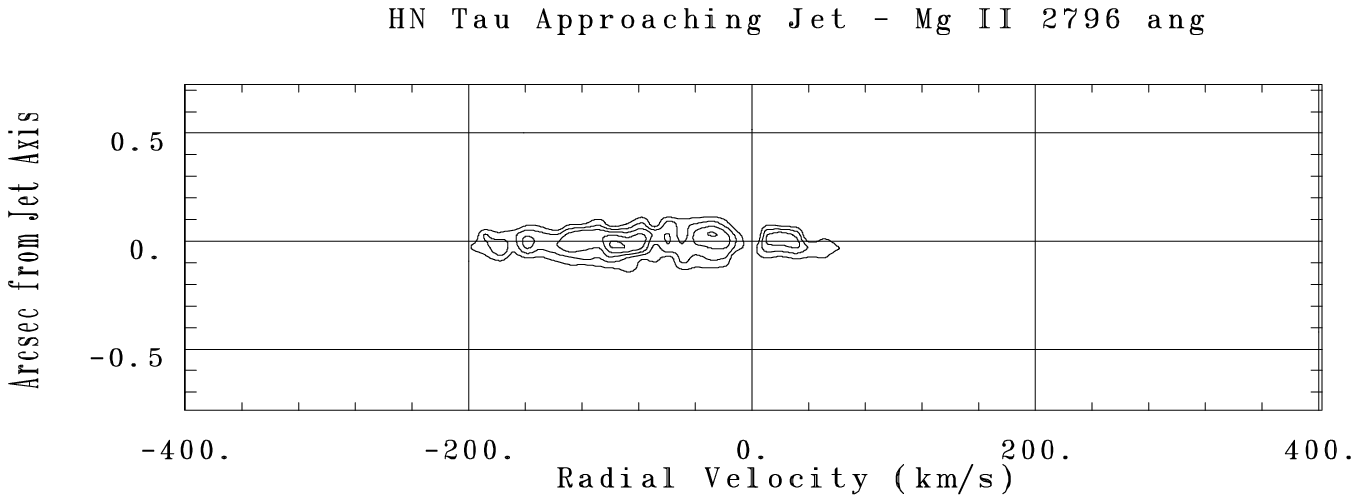}
\plotone{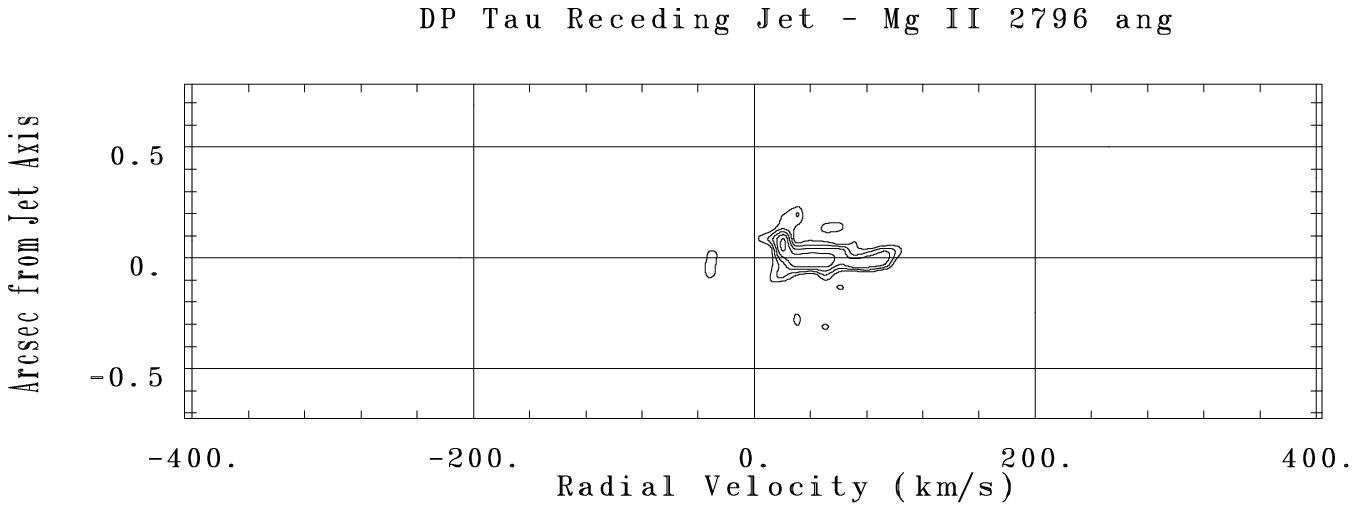}
\caption{Position-velocity contour plots in \ion{Mg}{2} $\lambda$2796 emission for all jet targets except that of CW Tau which was too faint to detect without binning the emission. Contour levels are given in Table~\ref{contours}. 
\label{pvs}}
\end{center}
\end{figure}

\begin{figure}
\begin{center}
\epsscale{1.0}
\plotone{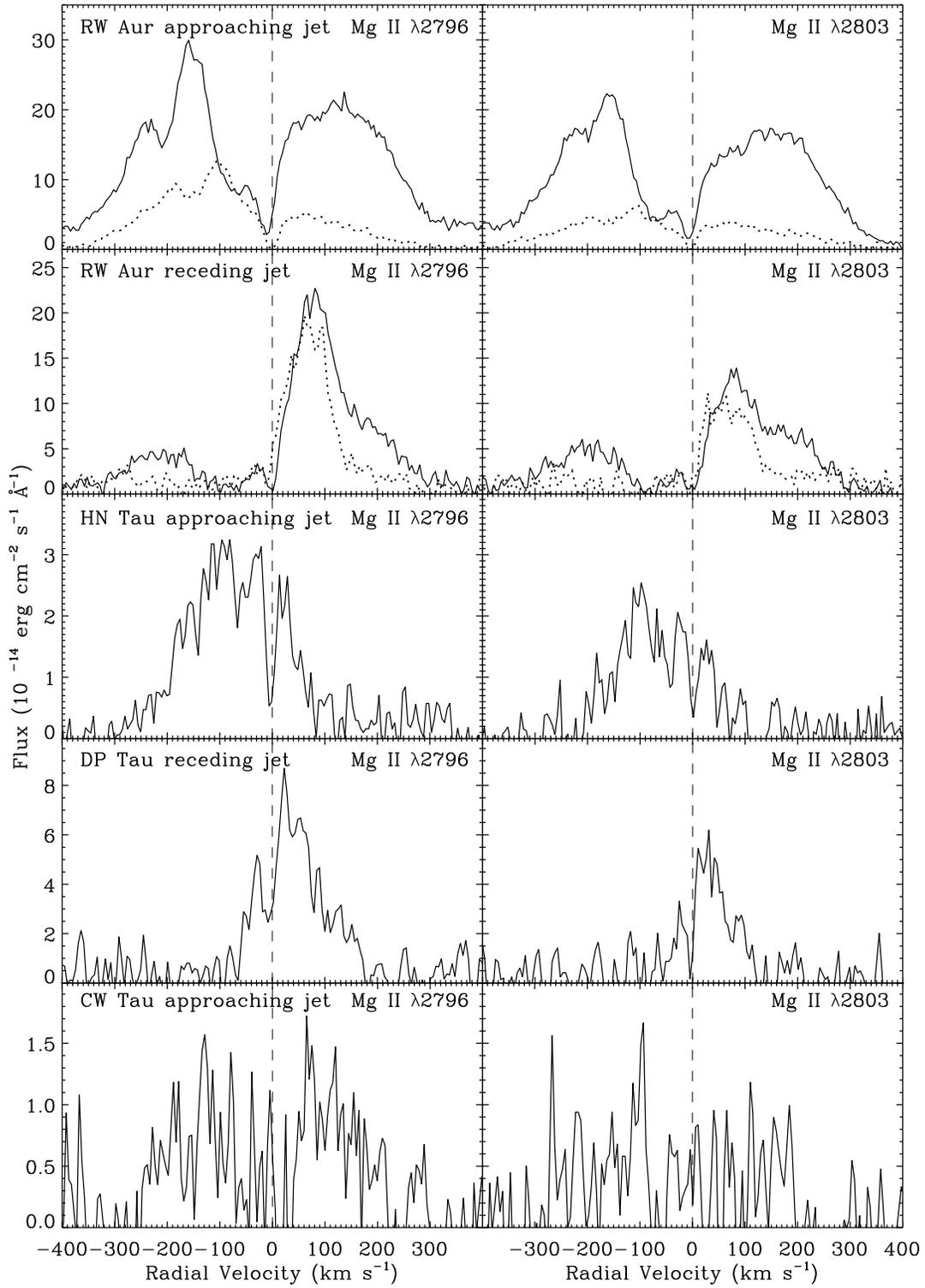}
\caption{\ion{Mg}{2} doublet in emission and absorption for several T Tauri jet targets observed close (within 50 AU) to the disk-plane. In the case of RW Aur, the dotted lines represents a second epoch (in an anti-parallel slit configuration) observed after a 6 month time interval. Note that with our chosen slit position, we escape the instrument PSF HWZM, but we are still affected by the diffraction rings since we see emission from the receding jet even when the slit is placed only over the approaching jet, and vice versa. The stellar continuum at this position is likely to be mostly absorbed by circumstellar material since it is only faintly observed here. 
\label{1dspec}}
\end{center}
\end{figure}

We have successfully detected the jet via emission of the NUV \ion{Mg}{2} $\lambda\lambda$2796,2803 doublet in all five jet targets. Position-velocity diagrams are shown in Figure~\ref{pvs} with corresponding contour levels in Table~\ref{contours}. Binning the spectra to 1D gives a clearer representation of the complexity of the spectral profile, especially in the case of the lower S/N targets, Figure \ref{1dspec}. Note that all velocities have been corrected to the systemic velocity, Table~\ref{v_r}. 

The S/N ratio is high in the case of the RW Aur bipolar jet, but not in the other targets. The reason for low S/N in the DP Tau, HN Tau and CW Tau spectra is possibly explained by the fact that these observations were conducted shortly after the identification of the MAMA detector dark current problem. However, the problem as also present during RW Aur observations, thus indicating it to be an intrinsically brighter target. 

In each spectrum, we see both blue- and red-shifted emission. With our observing mode, the slit is placed at either 0$\farcs$2 or 0$\farcs$3 along the jet and so it does not cover the star. At this position, we escape the instrument spatial point spread function (PSF) half-width at zero maximum (HWZM) of 0$\farcs$145, but we are still affected by the diffraction rings of the PSF. Therefore, we see emission from the receding jet even when the slit is placed over the approaching jet only, and vice versa. Nevertheless, this does not present a problem since the high spectral dispersion separates the emission to ensure we can examine each jet lobe individually. 

\subsection{\ion{Mg}{2} absorption}

There are a number of absorption features which become immediately evident upon binning, Figures \ref{1dspec}. See Table~\ref{absorption} for parameters. The very low velocity blue-shifted absorption dip at $\sim$\,-10\,km\,s$^{-1}$ features in all spectra, including those previously published as the first phase of this study \citep{Coffey2007}. A similar feature is identified in several T Tauri star spectra and is usually attributed to absorption by interstellar clouds, but is more likely to also include a contribution from non-LTE processes in the T Tauri system itself based on larger than expected equivalent widths and full width half maxima (FWHM), as discussed in detail by \citet{Ardila2002}. A similar claim can be made for the low velocity blue-shifted absorption dip here since we measure values in the same range (Table~\ref{absorption}). 

There is also a clear broad blue-shifted absorption trough extending from $\sim$\,-50 to -150\,km\,s$^{-1}$ in the higher S/N case of the RW Aur receding jet spectrum. This is the typical profile associated with an absorbing wind, and demonstrates that the blue-shifted outflow is seen in both emission and absorption. This feature is also present in the approaching jet spectrum, but does not reach as deep due to the higher flux. 

\begin{figure}
\begin{center}
\plottwo{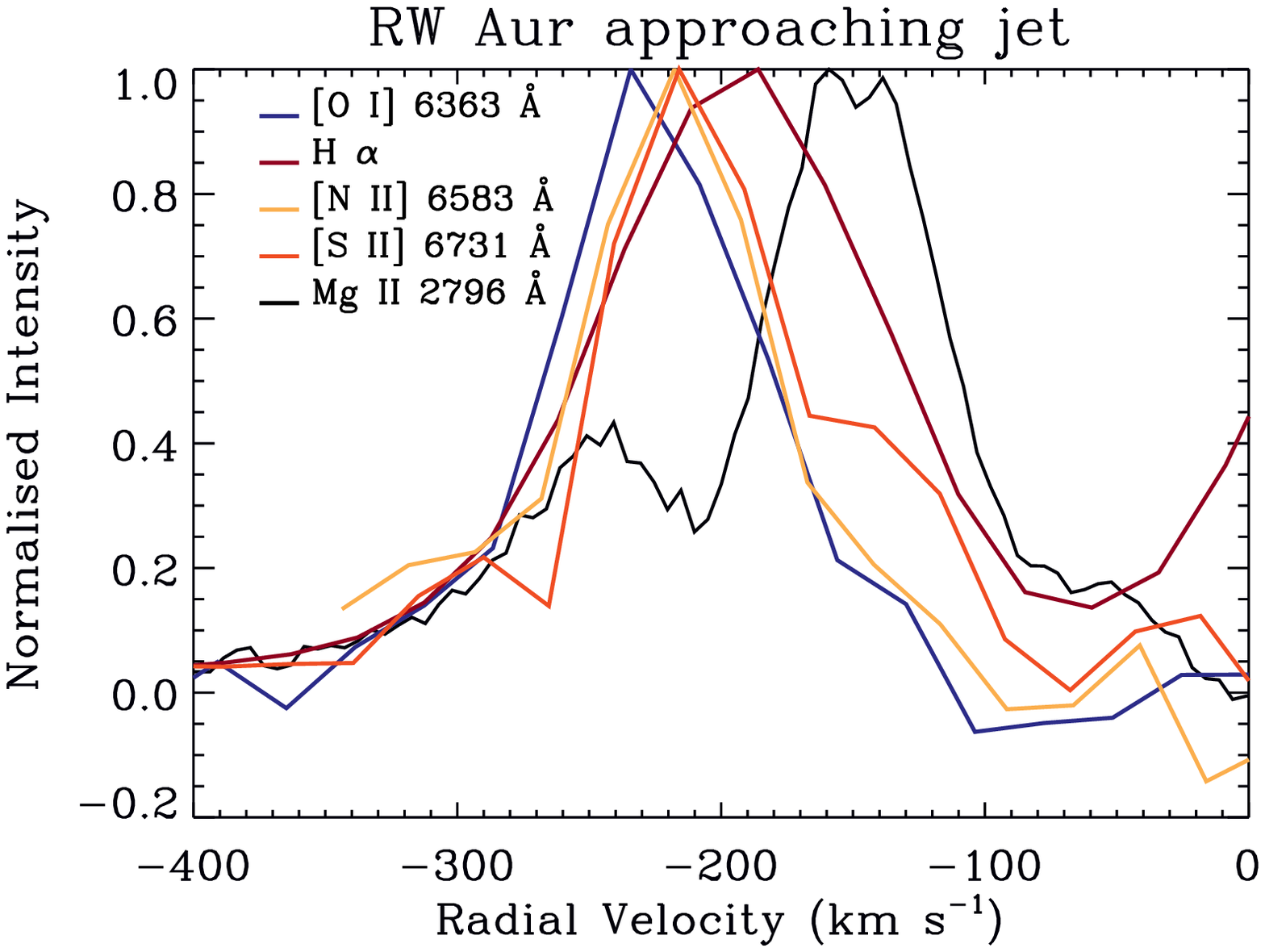}{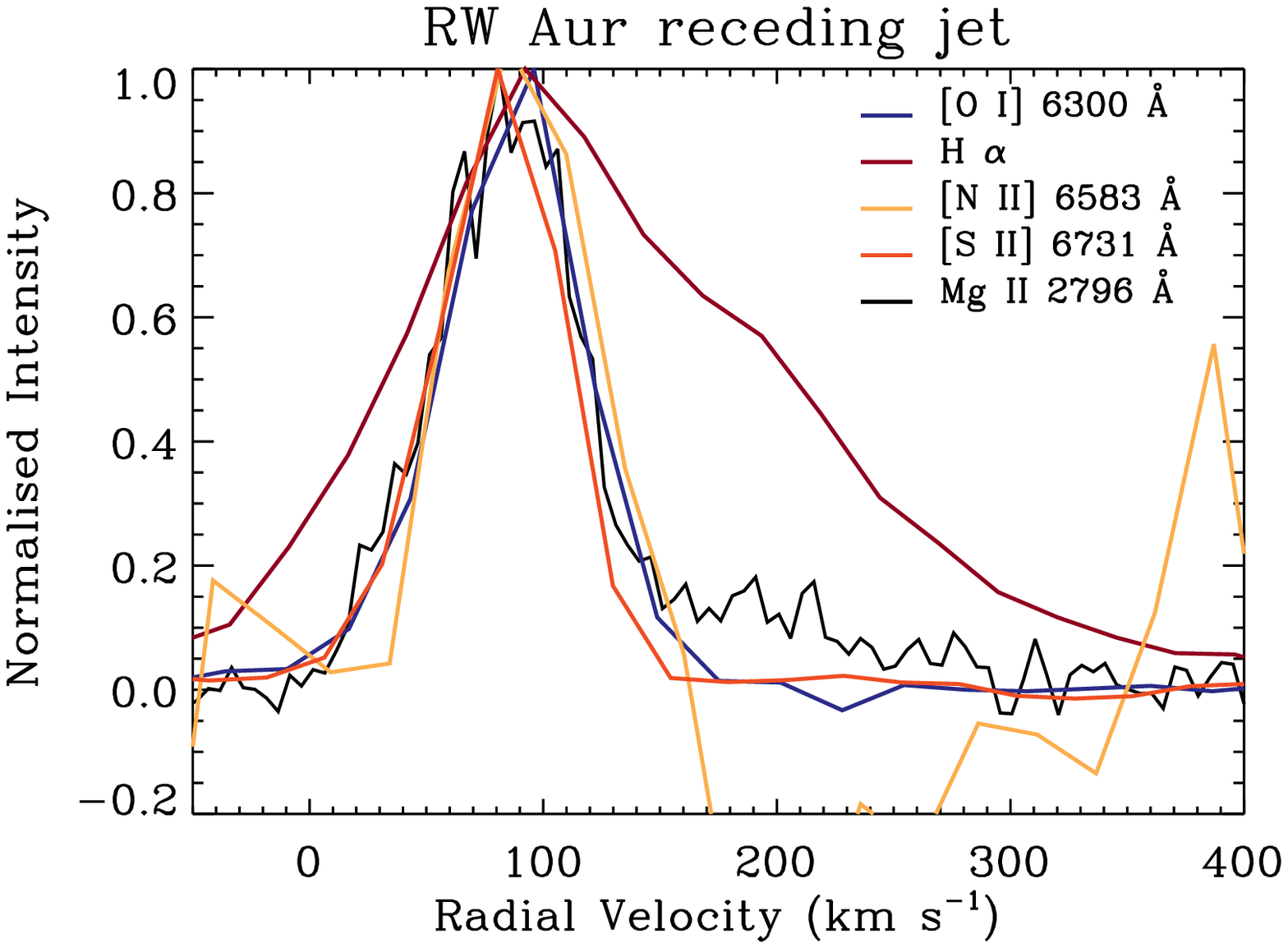}
\plottwo{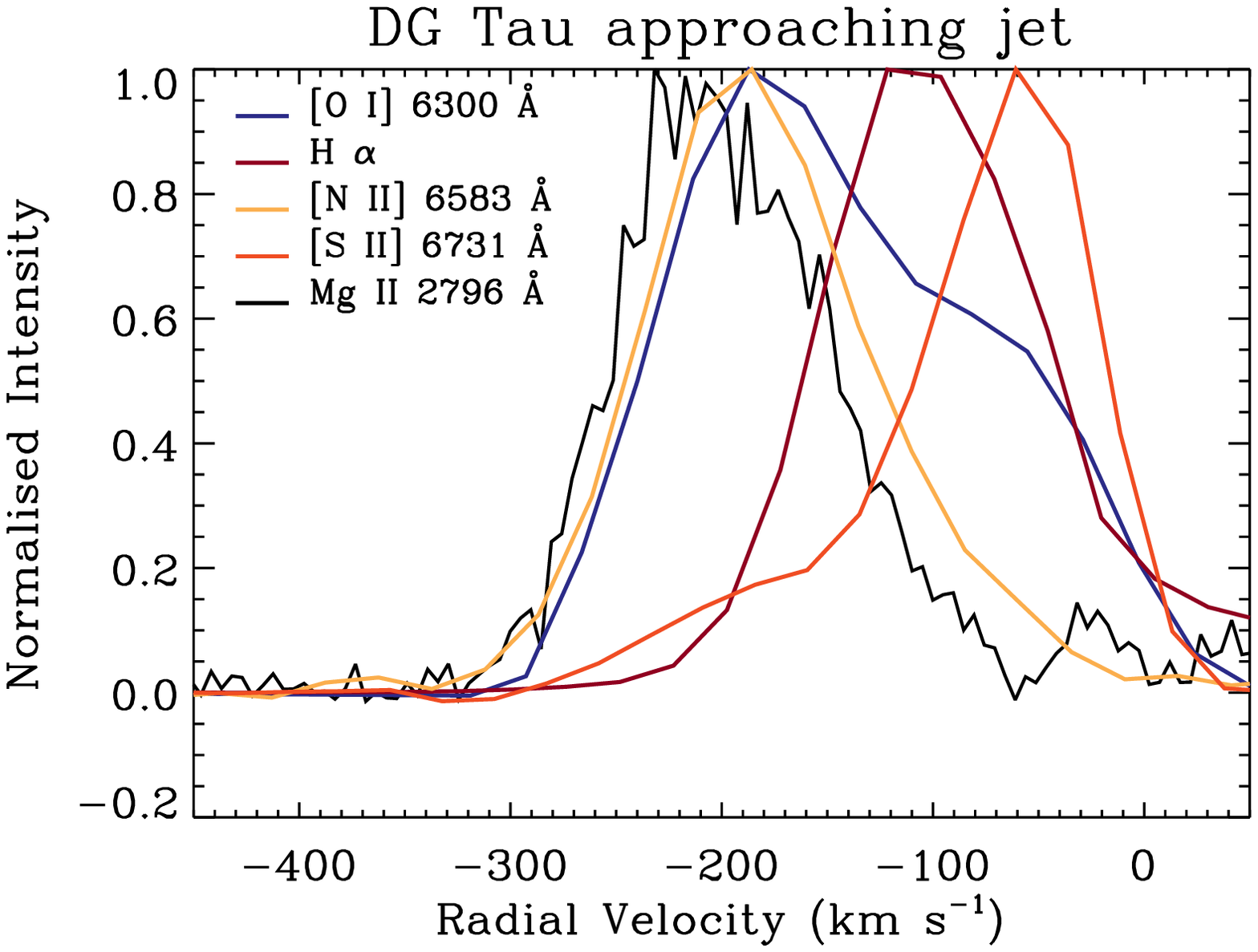}{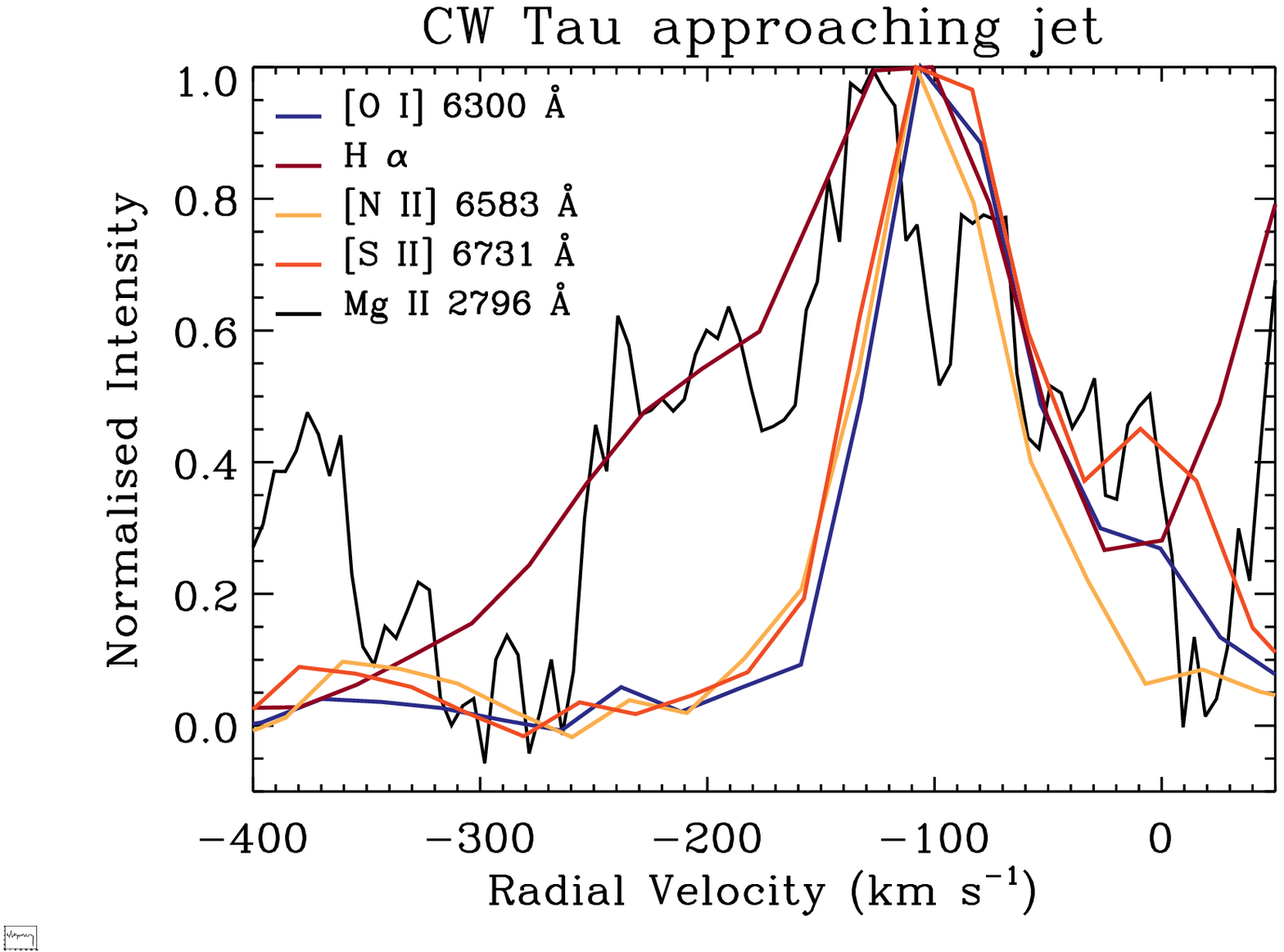}
\plottwo{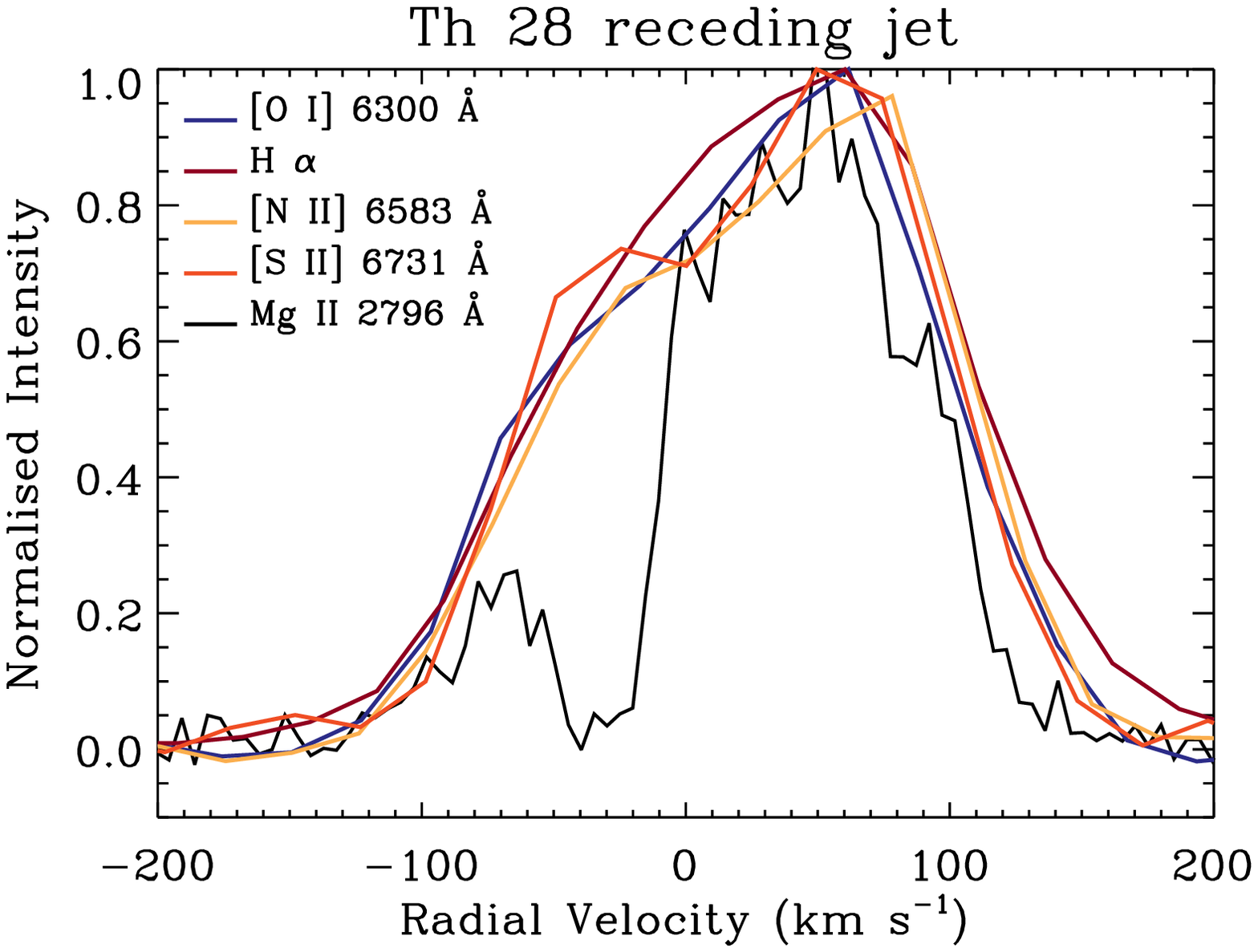}{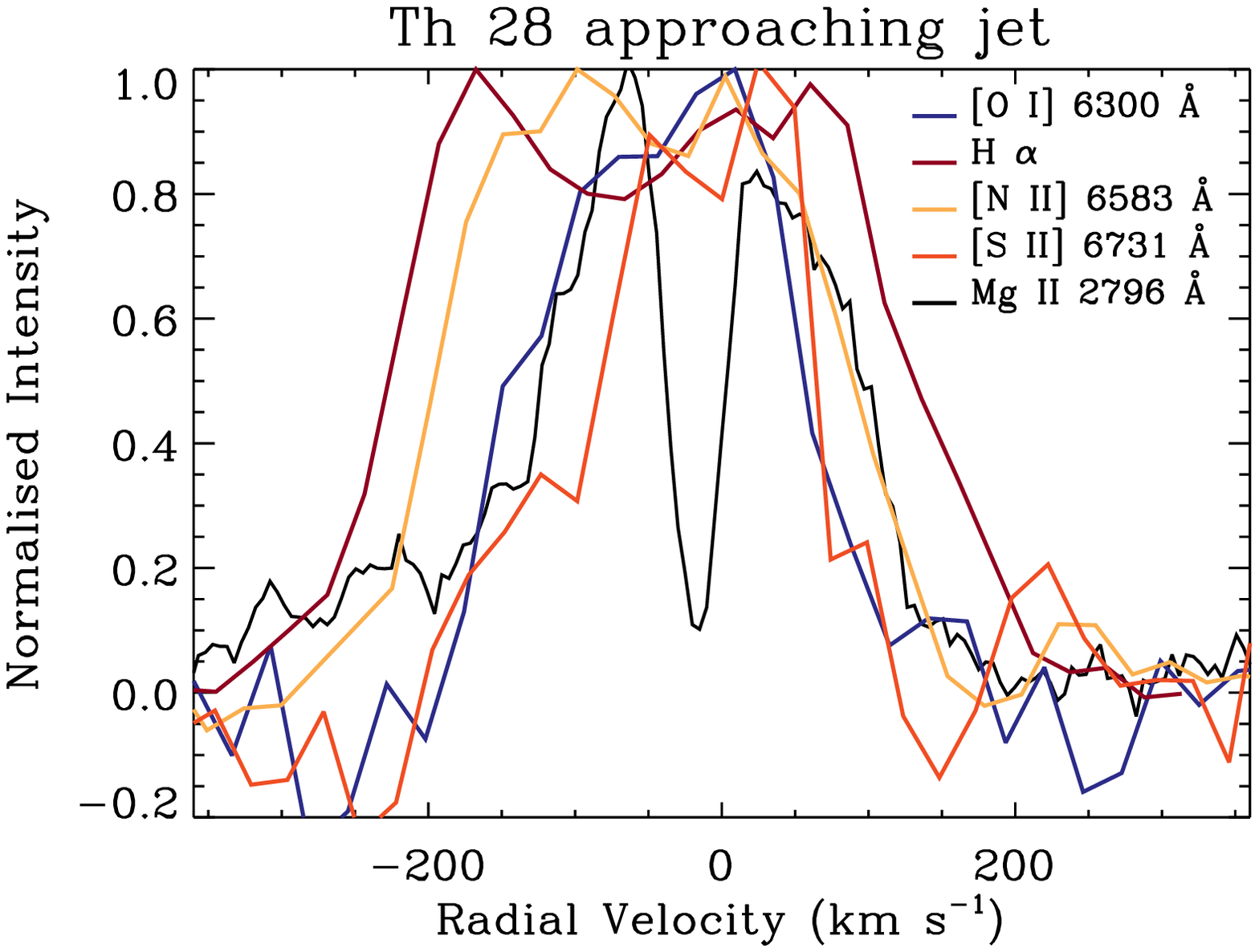}
\caption{Multi-wavelength picture of T Tauri jets close to their source ($<$ 100 AU). Emission on-axis is plotted, rather than binning all emission which can drown out features. However, due to low S/N, binning and smoothing was conducted for Th 28 [\ion{S}{2}] and \ion{Mg}{2} emission and CW Tau \ion{Mg}{2} emission. 
In all cases, there is emission from the other lobe of the bipolar jet but since there is an obvious dip at zero dividing the two emission components only one is shown, except for Th\,28 where there is no dip at zero.  
\label{multiwavelength}}
\end{center}
\end{figure}

Lastly, we can identify a third feature: a narrow high velocity blue-shifted absorption dip at $\sim$\,-~210\,km\,s$^{-1}$ apparent in the RW Aur approaching jet. Overlapping with our previously published optical HST/STIS spectra obtained at the same position (\citealp{Coffey2004}; 2007), Figure~\ref{multiwavelength}, we see that the optical forbidden emission lines peak at the same velocity as the dip in the \ion{Mg}{2} permitted line. Therefore, the jet is absorbing at these high velocities. We have also included plots for the DG\,Tau approaching jet and the Th\,28 bipolar jet from our previous optical and NUV observation, which nicely illustrate the coincidence of emission and absorption velocities within the plasma. 

\subsection{Shock-excited emission} 

Along with the strong \ion{Mg}{2} doublet, we have also identified several other NUV jet emission lines in the brightest spectra, i.e. those of the RW Aur approaching jet. Table~\ref{fluxes_nuv} lists emission lines, fluxes, S/N and FWHM for each target. Flux measurements were made after spectral binning, by  fitting the peak spatially with a Lorentzian profile which was a more accurate fit of the wings and width of the spatial profile. S/N was obtained by dividing the flux by the product of the Lorentzian FWHM and the square-root of the rms noise. Although strictly speaking this calculation is mathematically valid only for Gaussian fitting, we consider the Lorentzian FWHM to yield a satisfactory first order approximation. For example, if we use a Gaussian (instead of Lorentzian) for RW Aur approaching lobe in \ion{Mg}{2} 2796 \AA, then Table\,\ref{fluxes_nuv} would read flux of 9.34 (12.91), FWHM of 0$\farcs$14 (0$\farcs$11) and S/N of 52 (58).

Comparison of the \ion{Mg}{2} emission with HST/STIS observations eight years earlier in 2002-2003 reveals the \ion{Mg}{2} NUV doublet to be one to two orders of magnitude stronger than optical forbidden emission lines (Tables~\ref{fluxes_opt} \& \ref{fluxes_nuv}). Furthermore, the comparison shows that the receding jet is bright in optical lines while the approaching jet is brighter in NUV lines. Meanwhile, in all our NUV spectra, S/N is very low for the emission lines of \ion{Fe}{2}, \ion{Si}{2}] and \ion{C}{2}]. We were guided in our identification by lines already reported in the bow shock and mach disk of HH 47 A with HST/FOS \citep{Hartigan1999} which, when compared with shock models, revealed the lines to be shock-excited with a large range of excitation conditions. 

Jet radial velocities were obtained via Gaussian fitting, Table~\ref{v_r}. Note that for the RW Aur approaching jet, the highest peak was fitted though it is not considered to give an accurate representation of the overall jet velocity since it suffers from considerable absorption. Values were in excellent agreement with lines in HST/STIS optical spectra obtained with the same instrument configuration in 2002-2003 and, for example, infrared Keck spectra taken in 2006 \citep{Hartigan2009}. The NUV emission traces higher velocities, comparable with [\ion{O}{1}] emission, rather than the lower velocities preferably traced by [\ion{S}{2}] lines. Indeed the 1D spectra show that aside from the velocity of the peak emission, there is a significant level of flux traveling at much higher velocities. For example, the RW Aur bipolar jet emission extends from -\,375\,km\,s$^{-1}$ to +\,300\,km\,s$^{-1}$, the HN Tau jet emission reaches -300\,km\,s$^{-1}$, the DP Tau jet reaches +175\,km\,s$^{-1}$ and lastly the CW Tau jet emission is detected to at least -250\,km\,s$^{-1}$, Figure~\ref{1dspec}. 

\subsection{Jet rotation} 

A signature of jet rotation should manifest itself as a difference in the radial velocity between gas on one side of the jet axis and the other. This requires the jet width to be spatially resolved. It also requires in practice that observations achieve sufficient S/N (i.e. $>$ 20) at the borders of the jet. Lastly, it requires sufficient velocity resolution to measure small differences across the jet. 

Differences in radial velocity across the jet were measured by cross-correlating pixel rows either side of the jet axis. (Gaussian fitting was not appropriate due to the complex spectral profile shape.) First the position of the jet axis was determined by spectrally binning the higher velocity emission which is more collimated and hence more accurately traces the jet axis. This binned emission was fitted with a Lorentzian curve to determine the position of the jet axis. (A Gaussian curve was not appropriate due to the presence of wings at the jet borders.) The original 2D spectra were then shifted so that the jet axis was centered on a given pixel row. In this way, pixel rows at equal distances either side of the jet axis could be cross-correlated to give velocity differences. 

\begin{figure}
\begin{center}
\plottwo{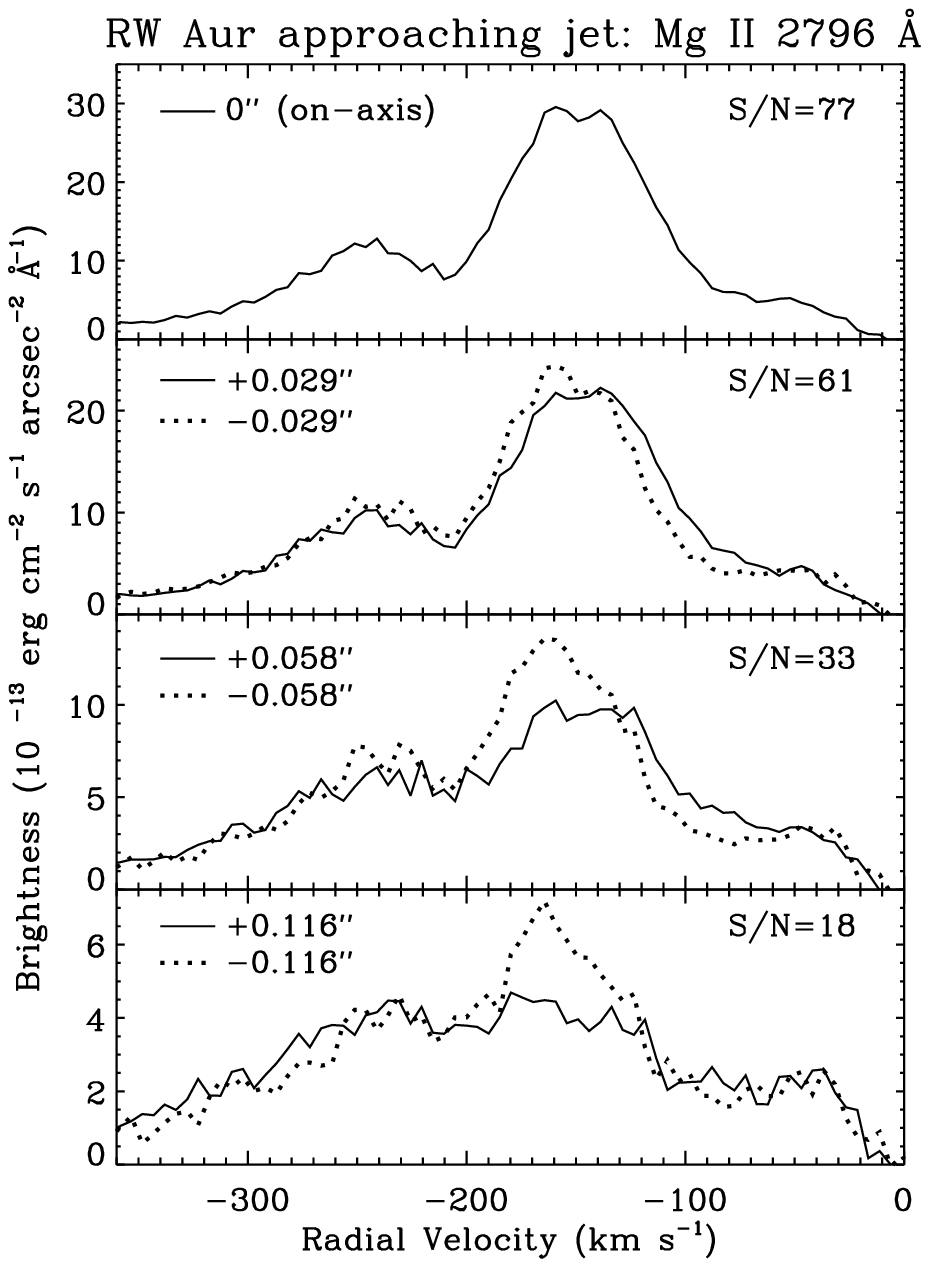}{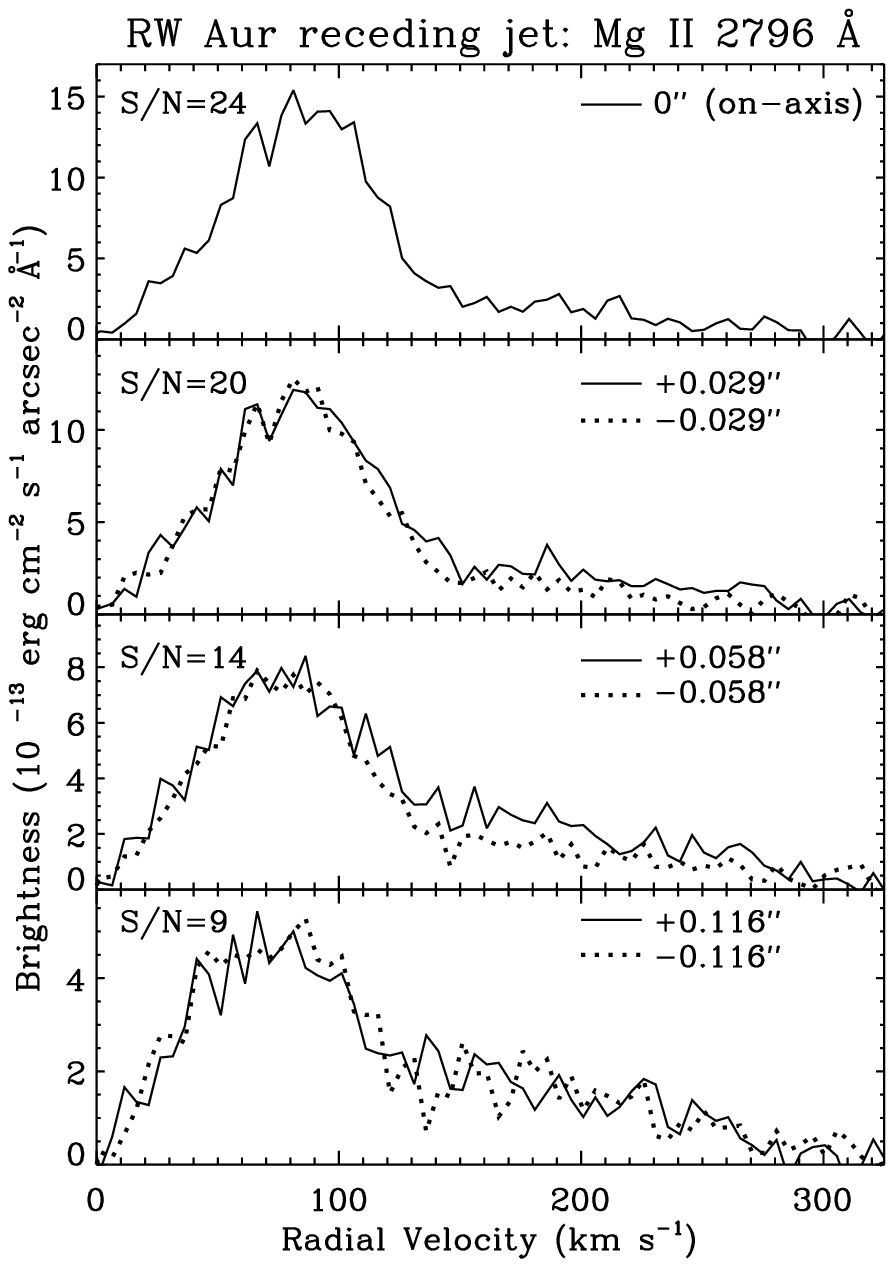}
\plottwo{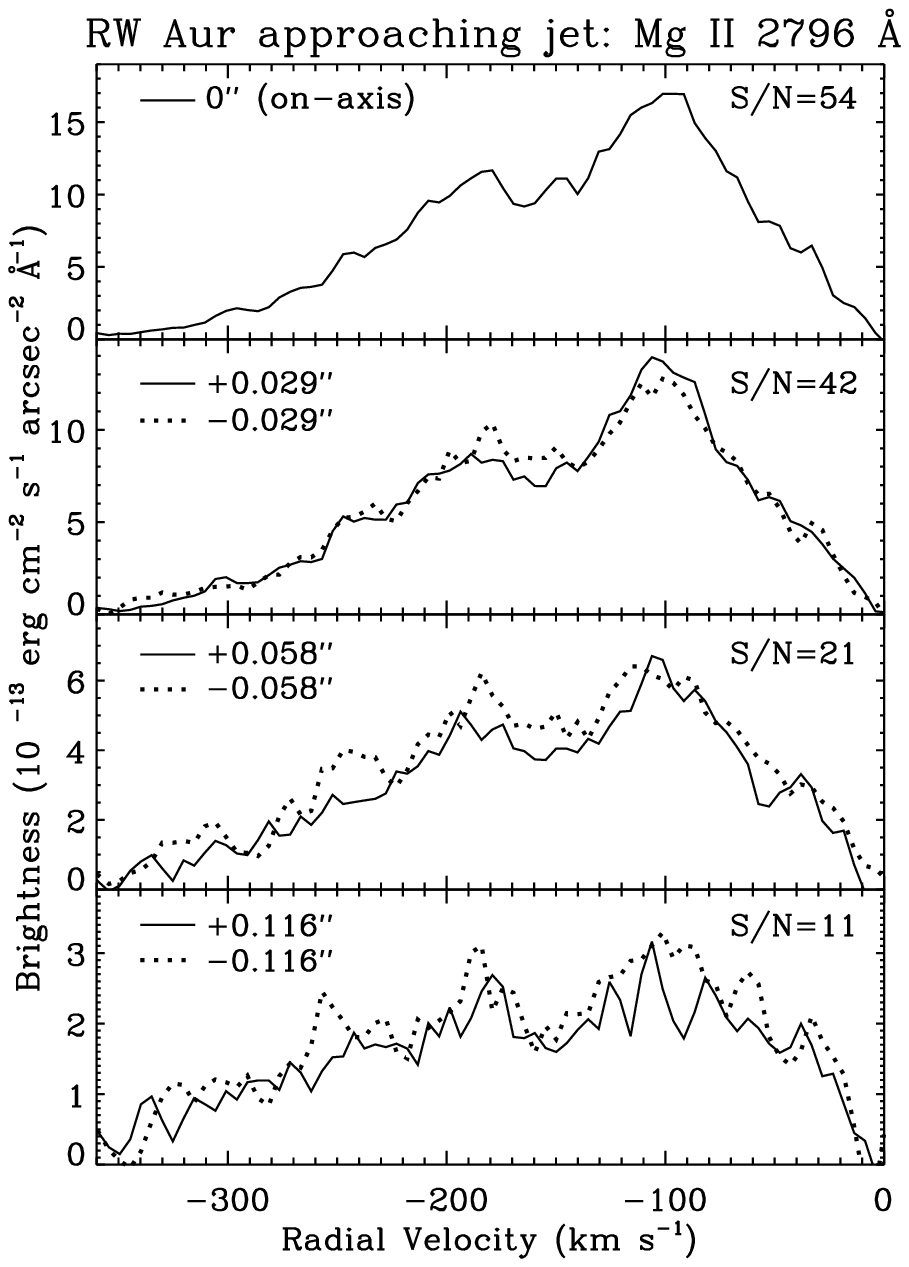}{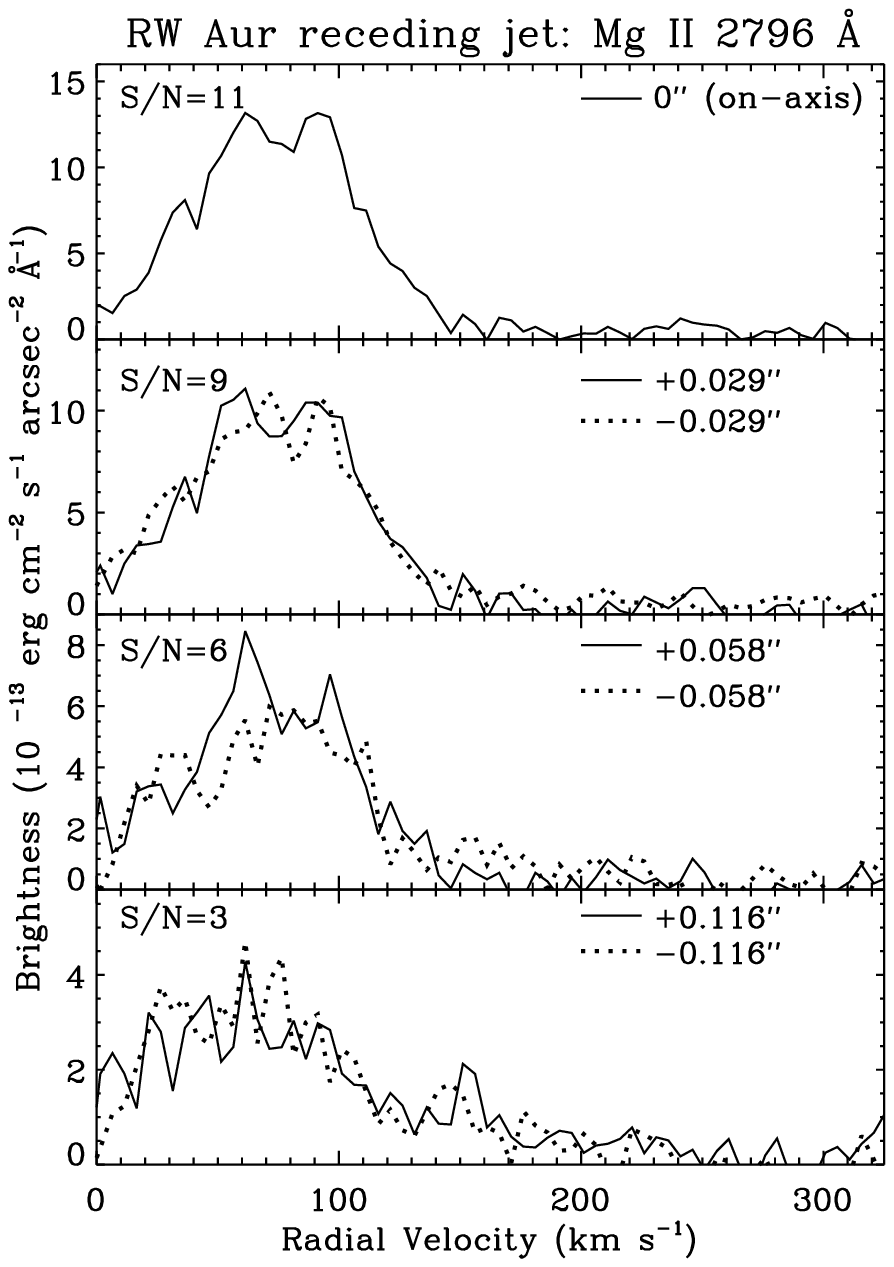}
\caption{Comparison of velocities either side of the jet axis for the RW Aur bipolar jet. The lower panel corresponds to anti-parallel slit observations taken after a 6 month time interval. A velocity difference of $\sim$ 10 km s$^{-1}$ is apparent in the first observations of the approaching jet at distances of $\pm$ 0$\farcs$029 and $\pm$ 0$\farcs$058. All other panels show not velocity differences. Note that the label $\pm$0$\farcs$029 indicates comparison of positions either side the jet axis, i.e. with a distance between them of 0$\farcs$058 equivalent to the resolution. 
\label{rotation_profiles}}
\end{center}
\end{figure}

RW Aur bipolar jet is the only target of sufficient S/N to allow this analysis. Comparison of the radial velocity profiles of gas equidistant either side of the jet axis is illustrated in Figure \ref{rotation_profiles}. It is clear in the first plot of the approaching jet that there is a difference in radial velocity across the jet in \ion{Mg}{2} $\lambda$2796 emission. A cross-correlation analysis reveals a difference of 7 km s$^{-1}$ at $\pm$ 0$\farcs$029 where S/N is 61, and 8 km s$^{-1}$ at $\pm$ 0$\farcs$058 where S/N is 33. Errors on the difference are heavily dependent on S/N. We achieve a 1 $\sigma$ error of 2.5 km s$^{-1}$ for S/N of 40, which increases to 5 km s$^{-1}$ for S/N of 20. As we continue to move away from the jet axis to $\pm$ 0$\farcs$116 and beyond, our S/N drops to below 20 and so differences are no longer identifiable. These results are reflected in the analysis of the second component of the doublet,  \ion{Mg}{2} $\lambda$2803, which presents a cross-correlation difference of 8 km s$^{-1}$ at $\pm$ 0$\farcs$029 where S/N is 43, and 2km s$^{-1}$ at $\pm$ 0$\farcs$058 where S/N is 28, after which S/N drops below 20. The receding jet was observed in the same night but shows no velocity difference. It is much fainter than the approaching jet with S/N of no more than 20 at $\pm$ 0$\farcs$029. 

Six months later, the anti-parallel slit data observation shows a drop in S/N, illustrating the \ion{Mg}{2} flux variability of this target. However, the S/N is still high enough at $\pm$ 0$\farcs$029 to reveal a gradient if one exists. A real signature (assuming it persists over time) should be inverted in an anti-parallel slit. Conversely, if measurements are due to instrumental effects, we will measure the same signature again. No difference in radial velocity is measured, and so we conclude that the original signature is not an artefact. It seems that the original signature is real, but that it has diminished or disappeared in the second epoch. The receding jet also shows a drop in S/N, and so again no difference in radial velocity is measurable across this jet. 

\section{Discussion} 
\label{Discussion} 

RW Aur is a well known variable system, which has been well studied due to its proximity and jet visibility close to the disk (see literature summary in  e.g. \citet{Hartigan2009}). Our observations confirm a strong asymmetry between the two lobes of the RW Aur bipolar jet. The approaching jet is traveling at twice the velocity of the receding jet, and is much brighter, in contrast to optical and near infrared spectra in which the receding jet was brighter in atomic lines (\citealp{Coffey2004}; \citealp{Woitas2005}; \citealp{Pyo2006}) and infrared atomic \citep{Hartigan2009} and molecular H$_2$ emission \citep{Beck2008}. This asymmetry in the RW Aur bipolar jet was first reported by \citet{Hirth1994} and is common among T Tauri bipolar jets, although the reason is as yet unknown. 

\subsection{Time variable ejection}

\begin{figure}
\begin{center}
\epsscale{1.0}
\plottwo{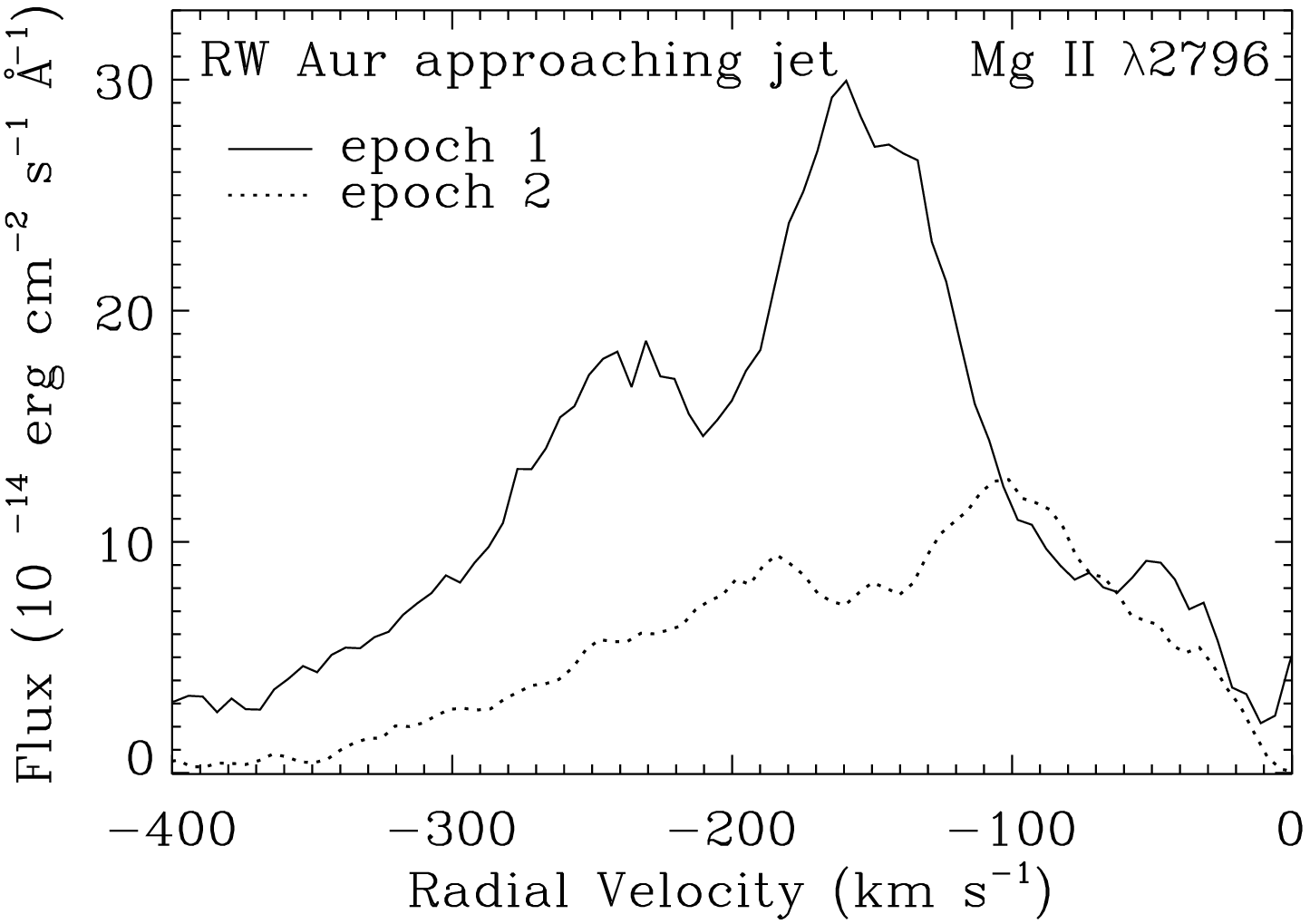}{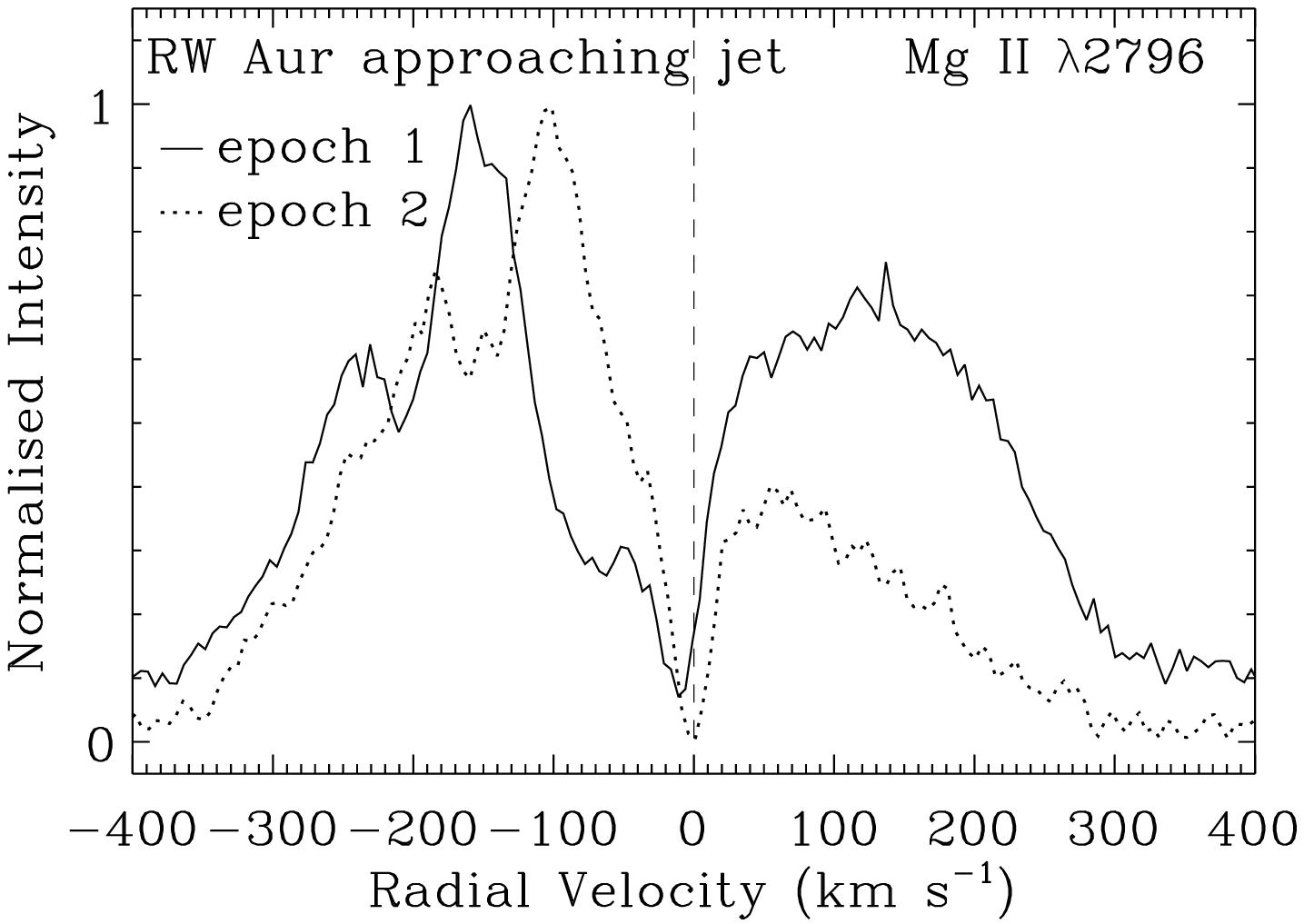}
\plottwo{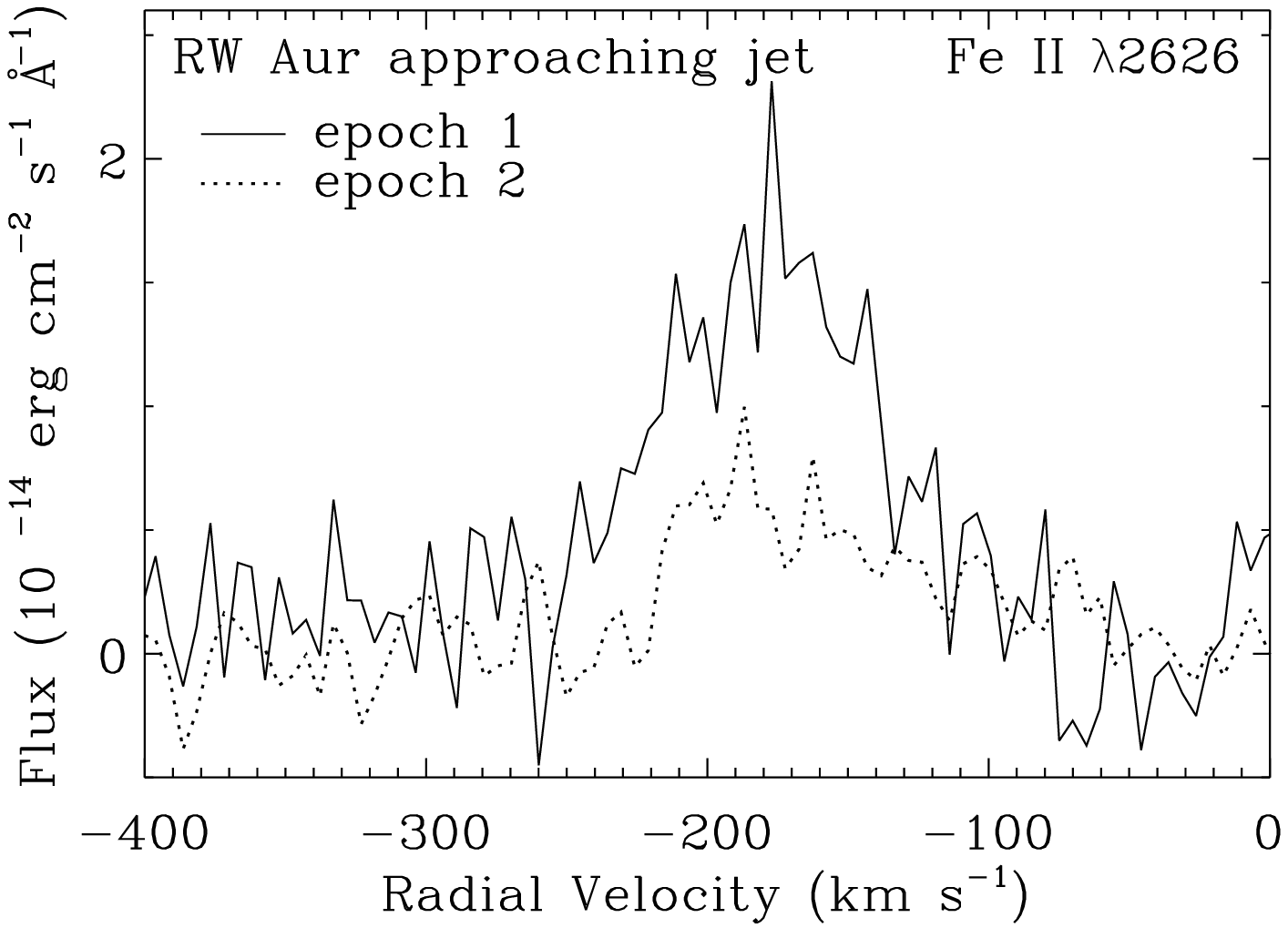}{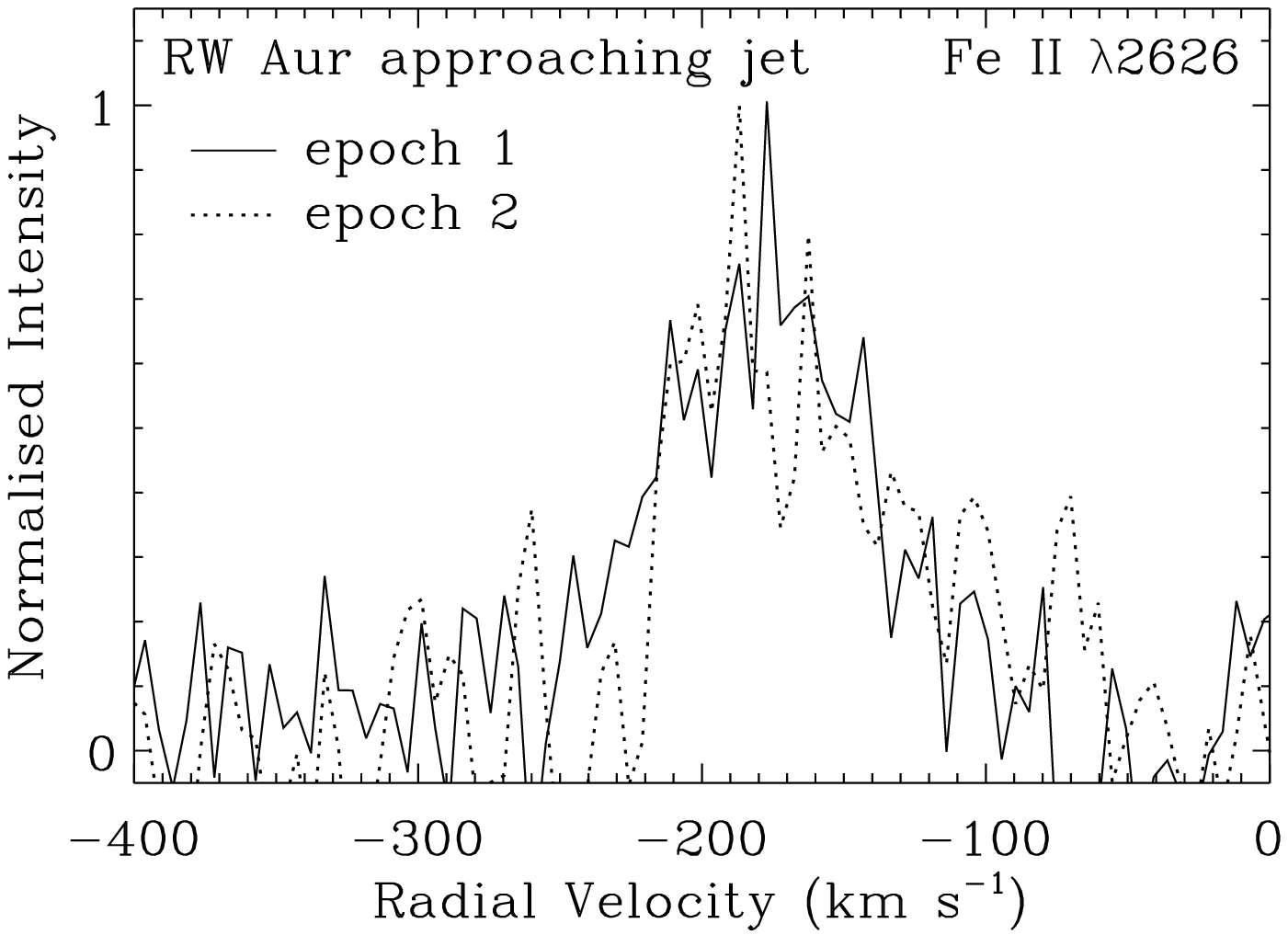}
\plottwo{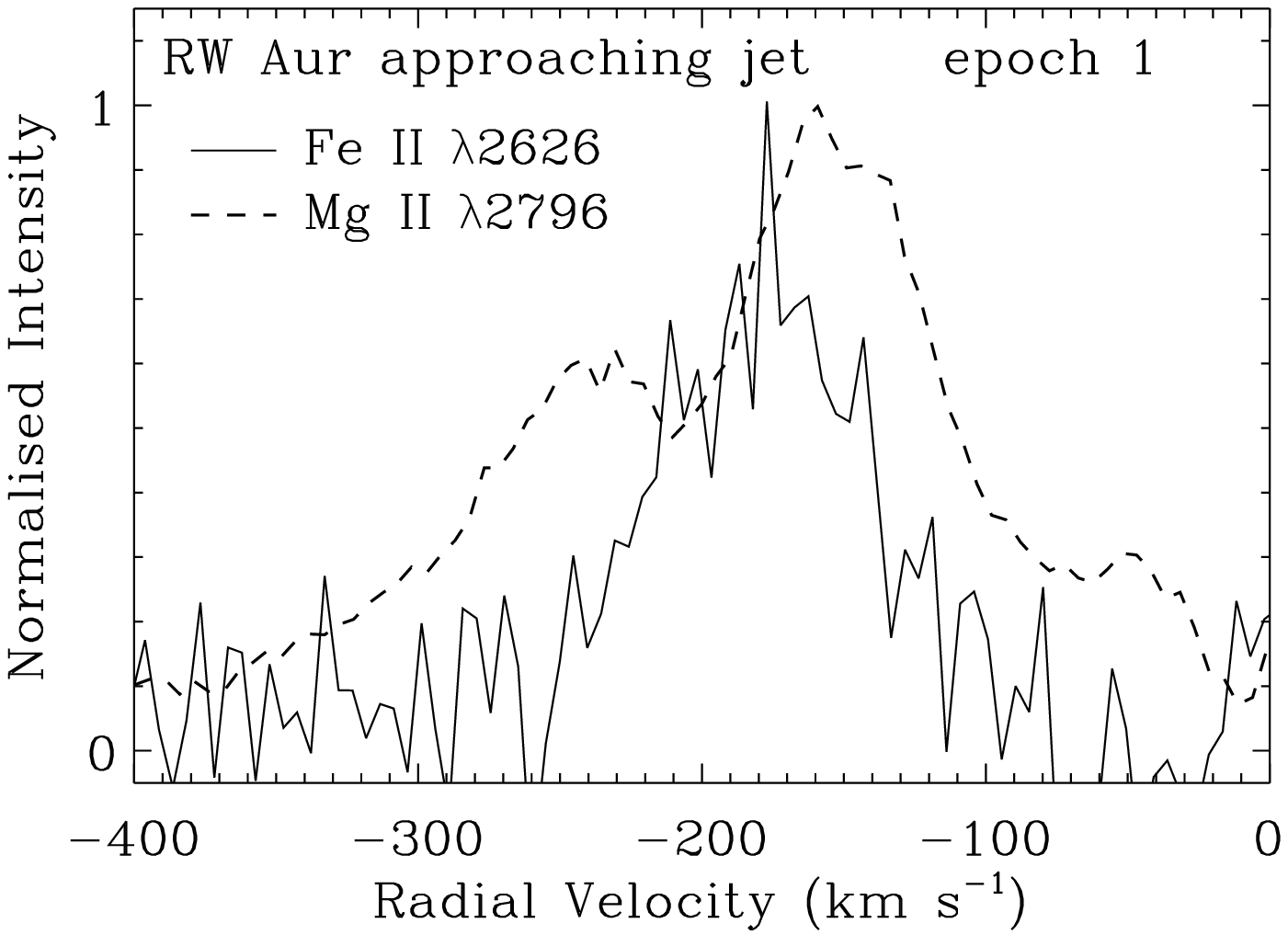}{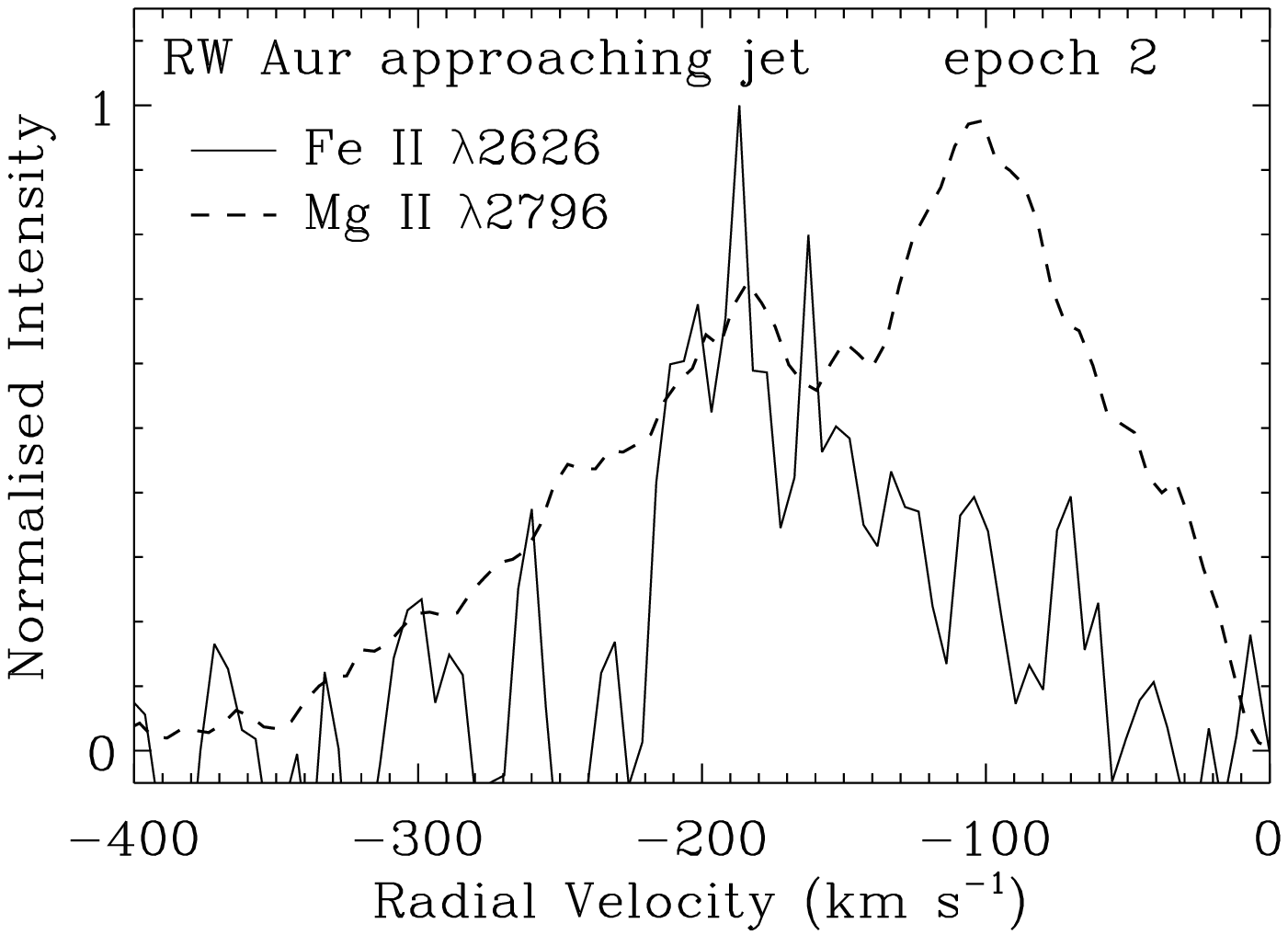}
\caption{RW Aur approaching jet in  \ion{Mg}{2} $\lambda$2796 and Fe II $\lambda$2626 emission, presented to highlight intensity and velocity differences in each line between epochs. 
\label{velprof}
}
\end{center}
\end{figure}

Our observations for RW Aur show that within a 6 month interval, both the flux and velocity of the bipolar jet have dropped significantly, Figure\,\ref{1dspec}. The approaching jet radial velocity decreases rapidly by $\sim$ 75 km\,s$^{-1}$, Table\,\ref{v_r}, which is a 35 \% drop assuming a jet velocity of 210 km\,s$^{-1}$ (i.e. that of the high velocity absorption which matches the position of the optical emission lines Figure~\ref{multiwavelength}). Meanwhile, the receding jet radial velocity decreases by $\sim$ 10 km\,s$^{-1}$, representing a 12 \% drop. Simultaneously, the approaching jet flux decreases by 42 \% while the receding jet flux decreases by 55 \%. Figure\,\ref{velprof}, emphasis the velocity and flux differences between the epochs. The drop in flux is mirrored in the NUV \ion{Fe}{2} $\lambda$2626 emission, though the velocity appears the same. Meanwhile, there is also a change in the FWHM of the receding jet between the two epochs, Table~\ref{fluxes_nuv}. 

We seem to be observing time variations in ejection. Assuming a jet radial velocity of 210 km\,s$^{-1}$, jet inclination to the plane of the sky of 44 $\pm$3 degrees \citep{L&oacute;pez-Mart&iacute;n2003}, and a distance to RW Aur of 139 $\pm$18~pc \citep{Bertout2006}, the time taken for the material to travel to the observed distance (0$\farcs$2) along the jet is 221 $\pm$ 37 days, while the interval between observations is 179 days (6 months). Clearly, we are observing variations in ejection on time scales of at least months. Indeed, the RW Aur system is well known for its variability \citep{Alencar2005}. Furthermore, the well known asymmetry between the two lobes appears to be borne out by the extent of ejection variability. 

\subsection{Angular momentum transfer} 

Only RW Aur's bipolar jet presents sufficient S/N for a jet rotation analysis. Our NUV data show that the approaching jet lobe has a gradient of $\sim$~10 km\,s$^{-1}$. Reassuringly, the direction matches the sense of disk rotation. A gradient is not found in the anti-parallel slit data taken six months later. However, there is a drop in jet radial velocity in the second epoch, so it is conceivable that any gradient caused by jet rotation has also dropped below detectable levels. The drop in flux, via a correlation between luminosity and density, also supports this since the toroidal field strength in the jet decreases with decreased mass loading as well as jet density (see e.g. \citealp{Pudritz2007}). 

Disconcertingly, this NUV gradient is opposite in direction to previously published RW Aur HST/STIS optical measurements which were interpreted as jet rotation (see Figure 1 of \citet{Coffey2004}). In the optical dataset, the receding jet presented the highest S/N, Table\,\ref{fluxes_opt}, while the approaching jet showed borderline S/N (when claiming whether a gradient is measurable) and so optical gradients for the approaching jet are not considered here. No NUV gradient is measured in the receding jet, again due to low S/N. 

The overall picture which has emerged for RW~Aur is this. In optical data 8 years ago, we see gradients in one direction in the receding lobe, but no gradient in the approaching lobe due to low S/N. Now, in NUV data, we see gradients in the opposite direction for the approaching lobe which disappear 6 months later, but no gradient in the receding jet due to low S/N. These results are not consistent with a simple jet rotation interpretation. However, we know of the inherent complexity and variability in the RW Aur system itself, and so any rotation signature is more likely to suffer contamination from factors external to the launching mechanism. Thus, our results lead us to conclude that this target is not suited to a jet rotation study. 

\section{Conclusion}

We successfully detect all five jet targets (from RW Aur, HN Tau, DP Tau and CW Tau) in several NUV emission lines, including \ion{Fe}{2}, \ion{Si}{2}], \ion{C}{2}] and the strong \ion{Mg}{2} doublet at 2800~\AA, in spectra obtained close to the base of the jet near the disk-plane where ejection takes place. These lines have been previously confirmed as arising in shocks and represent a range of excitation conditions. Several \ion{Mg}{2} absorption features are also present and identified as arising from the high velocity component of the jet and from non-LTE processes in the T Tauri system itself. 

S/N was low in all lines except the \ion{Mg}{2} doublet in the bipolar jet from RW Aur, which was observed in 2 epochs. The asymmetry in jet velocities is confirmed with the approaching jet traveling at about twice the speed of the receding jet. Furthermore, the approaching jet is brighter than the receding jet in NUV lines, although this is contrary to optical data taken at the same position eight years previously. RW Aur is know to be a highly variable system, and here we report variable ejection on timescales of months. The outflow flux drops by 42 and 55 \%, in the approaching and receding jet respectively, and outflow radial velocity drops by 35\% and 12 \% (i.e. 75 and 10 km\,s$^{-1}$). The well known asymmetry in these jet lobes is borne out by their asymmetric change in ejection activity. The time between the two observations is six months, though the emission has taken at least 7.5 months (assuming steady velocity) to arrive at the deprojected distance along the jet of 28~AU from the disk-plane. 

Our jet rotation analysis reveals a gradient of 10 km\,s$^{-1}$ across the RW Aur approaching jet at 0$\farcs$2 from the disk-plane. The direction of the gradient matches that of the disk. However, it is opposite in direction to the previously reported optical gradient of 25 km\,s$^{-1}$ in the receding jet at 0$\farcs$3 from the disk-plane. No NUV gradient is measured in the approaching jet 6 months later, but the jet radial velocity has dropped (by 75 km\,s$^{-1}$) which would cause a drop in toroidal velocity and so the gradient may not be measurable. No NUV gradient is observed in the receding jet in either epoch, but in these cases the S/N is less than 20 and so gradients are not detectable. These measurements are not consistent with a simple jet rotation interpretation. However, the complexity and variability of the RW Aur system may be a significant contributor to disruption of a signature of rotation. Our results highlight this difficulty, which renders RW Aur unsuitable for jet rotation studies. 

Although our results in this case are not consistent with a rotation signature, they cannot be assumed to undermine the validity of jet rotation observations in other systems, and thus the overall validity of magneto-centrifugal ejection models. Searching for jet rotation is a very difficult and slow program, since it pushes the limits of current instrumentation, but the return is potentially great. Clearly more high resolution observations are needed close to the jet base in other systems to build observational support for magneto-centrifugal models in which jets extract angular momentum from newly forming stars. 
\\ 

\acknowledgments

Based on observations made with the NASA/ESA Hubble Space Telescope, obtained
at the Space Telescope Science Institute, which is operated by the Association of Universities for Research in Astronomy, Inc.,
under NASA contract NAS 5-26555. These observations are associated with program 11660. 
D.C., E.R and F.B. acknowledge financial contribution under Italian Space Agency agreement ASI-INAF I/009/10/0. 
D.C. also wishes to acknowledge funding received from the Irish Research Council for Science, Engineering and Technology and European Community's Marie Curie European Reintegration Grant, under agreement FP7-MC-ERG-2008-239350. We gratefully acknowledge the STSci team, in particular Phil Hodge, for all the helpdesk support in ensuring the most accurate data recalibration. 

{\it Facilities:} \facility{HST (STIS)}


\begin{table}
\begin{center}
\caption{Details of observations \label{exposure_times}}
\begin{tabular}{lllccll}
\tableline\tableline
Target			&Jet PA		&Slit PA\tablenotemark{a}	&Observation		&No. of	&Exposure time		\\
				&(degrees)	&(degrees)				&date			&orbits	&(s)					\\  
\tableline
RW Aur approaching lobe	&130 	&265						&2010 Aug 20		&3		&2449 + (2983 x 2) 		\\
RW Aur receding lobe	&130 	&265						&2010 Aug 20		&2		&2983 $\times$ 2 		\\
HN Tau approaching lobe	&171 	&304\tablenotemark{b} 			&2010 Nov 18		&3		&2190 + (2728 $\times$ 2) \\
DP Tau receding lobe	&40 		&175 						&2010 Dec  ~3		&3		&2282 $\times$ 3	 	\\
CW Tau approaching lobe	&151 	&110\tablenotemark{b} 			&2010 Dec 27		&3		&2500 $\times$ 3	 	\\
RW Aur approaching lobe	&130 	&85\tablenotemark{c}			&2011 Feb 26		&3		&2449 + (2983 $\times$ 2) \\
RW Aur receding lobe	&130 	&85\tablenotemark{c}			&2011 Feb 26	 	&1		&2983 				\\
\tableline
\end{tabular}
\tablenotetext{a}{The Slit PA parameter is obtained by adding 135$^{\degr}$ to the jet PA, thus physically positioning the slit perpendicular to the direction of jet propagation.}
\tablenotetext{b}{HN Tau and CW Tau have slit positions misaligned by 2$^{\degr}$ and 4$^{\degr}$ respectively, due to restrictions on guide star availability.}
\tablenotetext{c}{Observations taken using anti-parallel slit (i.e. slit turned 180$^{\degr}$ with respect to previous observations of this target).}
\end{center}
\end{table}

\begin{table}
\begin{center}
\caption{Contour levels for Figure~\ref{pvs} \label{contours}}
\begin{tabular}{llllc}
\tableline\tableline
Target			&Floor	&Ceiling		&Interval		&Number \\
				&		&			&			&		\\
\tableline
RW Aur approaching lobe	&1.0		&29.2		&3.1			&10		\\
RW Aur receding lobe	&1.0		&13.7		&1.4			&10		\\
HN Tau approaching lobe	&1.0		&2.9			&0.5			&5		\\
DP Tau receding lobe	&1.0		&1.8			&0.2			&5		\\
\tableline
\end{tabular}
\tablecomments{Contour levels are on a linear scale in units of 10$^{-13}$ erg cm$^{-2}$ s$^{-1}$ \AA$^{-1}$ arcsec$^{-2}$}. 
\end{center}
\end{table}

\begin{table} 
\begin{center} 
\caption{Jet radial velocities. \label{v_r}} 
\begin{tabular}{lcccc} 
\tableline\tableline 
Target				&$v_{sys}$		&{$v_r$} 					&{$v_r$}		&{$v_r$}				\\
					&				&in \ion{Mg}{2} 			&in [\ion{O}{1}] 	&in [\ion{S}{2}]			\\ 
					&(km\,s$^{-1}$)	&(km\,s$^{-1}$) 			&(km\,s$^{-1}$)&(km\,s$^{-1}$)		\\  
\tableline 
RW Aur approaching jet (epoch 1)	&+23.5	&-152\tablenotemark{b}		&-180 to -200 & -120	\\
RW Aur receding jet (epoch 1)		&+23.5	&+85					&+105 to +115 &+100 	\\
HN Tau approaching jet			&+17.5\tablenotemark{a} &-83			&... &...				\\
DP Tau receding jet				&+17.5\tablenotemark{a}	&+42		&... &...				\\
CW Tau approaching jet			&+14.5	&-130					&-100 to -115 &-98		\\
RW Aur approaching jet (epoch 2)	&+23.5	&-101\tablenotemark{b}		&-180 to -200 &-120		\\
RW Aur receding jet (epoch 2)		&+23.5	&+74					&+105 to +115 &+100	\\
\tableline 
\end{tabular} 
\tablenotetext{a}{The systemic velocity for HN Tau and DP Tau is unknown, but an estimate of +17.5 km\,s$^{-1}$ for DP Tau \citep{Herczeg2007} is adopted and also applied to HN Tau as a reasonable approximation given they are in the same cloud.}
\tablenotetext{b}{This is the velocity of the \ion{Mg}{2} $\lambda$2796 emission peak after significant absorption and thus is {\em not} an accurate representation of the overall jet velocity in the near-UV, but is nevertheless useful to demonstrate jet deceleration between epochs. In other targets, the absorption does not significantly disrupt the peak and so velocities are a more accurate representation.} 
\tablecomments{Jet radial velocities, $v_r$, in near-UV \ion{Mg}{2}$\lambda$2796 emission compared with optical [\ion{O}{1}] $\lambda$6300,6363 and [\ion{S}{2}] $\lambda$6718,6732 emission in HST/STIS spectra obtained with the same instrument configuration about eight years earlier in 2002 and 2003. 
Values are corrected for the systemic velocity, $v_{sys}$, taken from \citet{Woitas2002} and \citet{Hartmann1986} for RW Aur and CW Tau respectively. }
\end{center}
\end{table}

\begin{table}
\begin{center}
\caption{Parameters of the \ion{Mg}{2} absorption features (Figure~\ref{1dspec}) \label{absorption}}
\begin{tabular}{lcccc}
\tableline\tableline
Frame \& feature		&\ion{Mg}{2} line	&Centre		&FWHM		&EW 	\\
					&\AA				&(km s$^{-1}$) &(km s$^{-1}$)	&\AA		\\
\tableline
RW Aur approaching jet	&				&			&			&		\\
~ dip 1				&2796			&-13			&25			&0.29	\\
					&2803			&-3			&20			&0.21	\\
~ dip 2				&2796			&-93			&51			&0.47	\\
					&2803			&-83			&67			&0.50	\\
~ dip 3				&2796			&-210		&36			&0.22	\\
					&2803			&-197		&34			&0.29	\\

RW Aur approaching jet (Epoch 2)	&		&		&			&		\\
~ dip 1				&2796			&-3		&28			&0.27	\\
					&2803			&-10		&43			&0.39	\\
~ dip 3				&2796			&-151	&40			&0.38	\\
					&2803			&-152	&48			&0.43	\\

RW Aur receding jet		&				&			&			&		\\
~ dip 1				&2796			&2			&20			&0.20	\\
					&2803			&-4			&26			&0.22	\\
~ dip 2				&2796			&-93			&89			&0.80	\\
					&2803			&-84			&67			&0.68	\\

HN Tau approaching jet	&				&			&			&		\\
					&2796			&-4			&17			&0.16	\\
					&2803			&2			&16			&0.15	\\

DP Tau receding jet		&				&			&			&		\\
					&2796			&-6			&26			&0.23	\\
					&2803			&-6			&14			&0.12	\\

\tableline
\end{tabular}
\tablecomments{Values are indicative only, since Gaussian fitting was approximate due to the fact that the underlying emission profile shape is unknown. Velocity centres have been corrected for the systemic velocity. There were no absorption parameters for epoch 2 of the RW Aur receding jet observation, because the blue-shifted emission in this frame is too faint to show up absorption.} 
\end{center} 
\end{table} 

\begin{table}
\begin{center}
\caption{Near-UV flux measurements of T Tauri jets close to the jet base.  \label{fluxes_nuv}}
\begin{tabular}{llllll}
\tableline\tableline
Target		&Emission line	&Flux							&FWHM	&FWHM	 		&S/N		\\
			&(\AA)		&(10$^{-14}$ erg cm$^{-2}$ s$^{-1}$)	 &(arcsec)	 &(\AA)			&		\\  
\tableline 
RW Aur blue lobe	&\ion{C}{2}] $\lambda$2324.21, 2325.4, 2326.11, 2327.64, 2328.83		&0.92	&0.11	&2.6		&7  \\ 
				&\ion{Si}{2}] $\lambda$2335.1		&0.38	&0.13		&0.7				&3  \\
				&\ion{Si}{2}] $\lambda$2344.92 + \ion{Fe}{2} $\lambda$2344.21 	&0.16	&0.10		&...				&2		 \\
				&\ion{Fe}{2} $\lambda$2382.76 	&0.22	&0.11	&...		&2	\\
				&\ion{Fe}{2} $\lambda$2612.65	&0.66	&0.18	&1.5		&8		\\
				&\ion{Fe}{2} $\lambda$2626.45	&0.52	&0.09	&0.8		&5		\\
				&\ion{Mg}{2} $\lambda$2796		&12.91	&0.11	&1.9		&58		\\
				&\ion{Mg}{2} $\lambda$2803		&10.75	&0.15	&1.7		&47		\\
RW Aur red lobe	&\ion{Mg}{2} $\lambda$2796	&8.60		&0.27	&1.2		&39				\\
				&\ion{Mg}{2} $\lambda$2803	&4.50		&0.29	&1.6		&20				\\
HN Tau blue lobe	&\ion{Mg}{2} $\lambda$2796	&2.03		&0.17	&1.4		&16				\\
				&\ion{Mg}{2} $\lambda$2803	&1.19		&0.16	&1.8		&9				\\
DP Tau red lobe	&\ion{Mg}{2} $\lambda$2796	&0.94		&0.23	&0.9		&8				\\
				&\ion{Mg}{2} $\lambda$2803	&0.64		&0.24	&1.0		&5				\\
CW Tau blue lobe	&\ion{Mg}{2} $\lambda$2796	&0.11		&0.08	&1.5		&2				\\
				&\ion{Mg}{2} $\lambda$2803	&0.07		&...		&...		&1				\\
RW Aur blue lobe\tablenotemark{a}	&\ion{C}{2}] $\lambda$2324.21, 2325.4, 2326.11, 2327.64, 2328.83		&0.29	&0.11	&...		&2  \\ 
				&\ion{Si}{2}] $\lambda$2335.1		&0.18	&0.11		&0.9				&3  \\
				&\ion{Si}{2}] $\lambda$2344.92 + \ion{Fe}{2} $\lambda$2344.21 	&0.15	&0.14		&...				&2		 \\
				&\ion{Fe}{2} $\lambda$2382.76 	&0.29	&0.15	&...		&2	\\
			 	&\ion{Fe}{2} $\lambda$2612.65	&0.33	&0.18	&...		&4	\\
 				&\ion{Fe}{2} $\lambda$2626.45	&0.13	&0.06	&1.1		&2	\\
				&\ion{Mg}{2} $\lambda$2796	&7.53	&0.09	&1.9		&62		\\
				&\ion{Mg}{2} $\lambda$2803	&3.95		&0.09	&2.0		&32				\\
RW Aur red lobe\tablenotemark{a} &\ion{Mg}{2} $\lambda$2796	&3.86	&0.13 	&0.9		&27		\\
				&\ion{Mg}{2} $\lambda$2803	&2.01		&0.13	&1.0		&16				\\
\tableline 
\end{tabular} 
\tablenotetext{a}{Second epoch observations taken after six months, and using a so-called anti-parallel slit configuration (i.e. the slit was turned by 180$\degr$ with respect to previous observations).} 
\end{center}
\end{table}

\begin{table}
\begin{center}
\caption{Optical flux measurements of T Tauri jets close to the jet base.  \label{fluxes_opt}}
\begin{tabular}{llllll}
\tableline\tableline
Target		&Emission line	&Flux		&FWHM	&FWHM	 		&S/N		\\
			&(\AA)		&(10$^{-14}$ erg cm$^{-2}$ s$^{-1}$)	 &(arcsec)	 &(\AA)	&		\\  
\tableline 
RW Aur blue lobe	&[\ion{O}{1}] $\lambda$6300 &0.15	&0.11	&2.6		&22 \\
				&[\ion{S}{2}] $\lambda$6731 &0.12	&0.12	&2.5		&18 \\
RW Aur red lobe	&[\ion{O}{1}] $\lambda$6300 &0.47	&0.14	&1.5		&69 \\
				&[\ion{S}{2}] $\lambda$6731 &0.44	&0.14	&1.4		&98 \\
CW Tau blue lobe	&[\ion{O}{1}] $\lambda$6300 &0.16	&0.24 	&3.4 	&20 \\
				&[\ion{S}{2}] $\lambda$6731 &0.08	&0.20	&3.9		&21 \\
\tableline 
\end{tabular} 
\tablecomments{Flux measurements from archival HST/STIS optical spectra obtained in 2002 and 2003 (proposal ID 9435) with the same instrument configuration as our NUV spectra.}
\end{center} 
\end{table} 






\end{document}